\documentclass[preprint,12pt]{elsarticle}

\usepackage{filecontents}
\usepackage{epstopdf}
\usepackage{subcaption}
\usepackage{tikz}
\usepackage{overpic}
\usepackage{import}
\usepackage{here}
\usepackage{pgf}
\usepackage[utf8]{inputenc}

\usepackage{booktabs}
\usepackage{enumitem}
\usepackage{geometry}
\usepackage{lineno}

\usepackage[version=3]{mhchem}
\usepackage{amsmath,amsthm,amssymb,amsfonts,amsbsy,txfonts}
\usepackage{mathrsfs}

\usepackage{bibspacing}

\usepackage{multicol}
\usepackage{nomencl}
\usepackage{etoolbox}

\usepackage{hyperref}

\usepackage{listings}
\usepackage{amssymb}

\journal{Computer \& Fluids}

\begin{document}

\begin{frontmatter}



\title{Dynamic load balance of chemical source term evaluation in high-fidelity combustion simulations}



\author{Guillem Ramirez-Miranda}
\author{Daniel Mira\corref{cor1}}
\cortext[cor1]{Corresponding author}
\ead{daniel.mira@bsc.es}
\author{Eduardo J. Pérez-Sánchez}
\author{Anurag Surapaneni}
\author{Ricard Borrell}
\author{Guillaume Houzeaux}
\author{Marta Garcia-Gasulla}
\address{Barcelona Supercomputing Center (BSC), Plaza Eusebi Güell 1-3, 08034, Barcelona Spain}

\begin{abstract}

This paper presents a load balancing strategy for reaction rate evaluation and chemistry integration in reacting flow simulations. The large disparity in scales during combustion introduces stiffness in the numerical integration of the PDEs and generates load imbalance during the parallel execution. The strategy is based on the use of the DLB library to redistribute the computing resources at node level, lending additional CPU-cores to higher loaded MPI processes. This approach does not require explicit data transfer and is activated automatically at runtime. Two chemistry descriptions, detailed and reduced, are evaluated on two different configurations: laminar counterflow flame and a turbulent swirl-stabilized flame. For single-node calculations, speedups of 2.3x and 7x are obtained for the detailed and reduced chemistry, respectively. Results  on  multi-node  runs  also show   that  DLB  improves  the  performance  of  the pure-MPI code similar to single node runs. It is shown DLB can get performance improvements in both detailed and reduced chemistry calculations.
\end{abstract}


\end{frontmatter}


\section{Introduction}
\label{}

Regulations applied to the power and transportation sectors have propelled research on optimization of thermal engines and other combustion systems, to simultaneously reduce fuel consumption and pollutant emissions. 
The extensive use of Computational Fluid Dynamics (CFD) for this purpose, as a conventional and indispensable tool, has made imperative the search of efficient algorithms that can solve the equations of chemically reacting flows at reduced computational costs. The chemical source terms are inserted in the transport equations, together with the evaluation of the convection and diffusion terms, that characterize the reacting flow.

Moreover, the choice of the reaction mechanism is a critical aspect when performing high-fidelity combustion simulations. The cost of the evaluation of the chemical source term depends on the size and stiffness of the reaction mechanism, and can result in the most demanding part of the calculation. While the 
workload
to evaluate the transport terms can be 
parallelized using standard domain decomposition strategies
\cite{Kodavasal_JERS_2016},
it is harder to define a computationally balanced distribution for the chemistry workload and this specific task can end up taking
90$\%$ of the computational time \cite{Shi2012, Zirwes2018}.
Such load imbalance can be explained by the nature of the combustion process, where the chemical reactions occur in specific regions of the domain and usually along thin layers, resulting, in consequence, in high disparity of computational load between processors, as not all the subdomains require the evaluation of the chemical reaction rates. Moreover, the stiffness of the ODEs (ordinary differential equations) system, caused by the non-linearity of chemistry and the wide range of time scales for radicals (around $10^{-8}$ s) and major species (in the order of ten of microseconds), increases the computational cost for the chemistry integration and accentuates even more the load imbalance.

However, different from the transport terms of the flow variables, which depend on the state at the vicinity at each point, chemical source terms only depend on the punctual thermochemical state of the mixture and, hence, it is susceptible to a high degree of parallelization. 

Based on these considerations, different strategies have been used to achieve load balance in combustion calculations. In the early works by Thevenin et al. \cite{Thevenin1996}, there was a transfer of points between neighboring processors, so that the nodes requiring more computing time than the average send a few grid points to their neighbors. A similar strategy was proposed by Antonelly and D'Ambra (2011) \cite{Antonelli2011}, where a cell distribution based on a dynamic load balancing that preserves contiguity of the computational grid cells was used.

Another important aspect is the stiffness of system, which can be noticeably different between cells. Chemical mechanisms may have between 50 and 1000 species, which include species featuring a wide range of time scales depending on the local conditions. For instance, the low temperature region determines the autoignition with high degree of stiffness, while the high temperature region has low stiffness as species with large mass fractions can reach equilibrium relatively fast~\cite{Antonelli2011}. These observations can be used to reduce the computational cost. Muela et al.~\cite{Muela2019} proposed a measure of the stiffness to separate between explicit and implicit algorithms. Kodavasal et al. (2016)~\cite{Kodavasal_JERS_2016} proposed a “stiffness-based” algorithm for load balancing chemical kinetics using 
information from previous time-steps.
A similar strategy has been used by Teckgül et al. (2021)~\cite{TEKGUL2021108073}, which also includes a zonal reference cell mapping method to avoid the evaluation of the kinetic rates in ambient regions with low reactivity. In the context of High Performance Computing (HPC) with massive use of CPUs, this strategy can be largely benefited by the hybrid use of CPUs and GPUs provided that explicit algorithms, which are easily parallelized and consume less memory, are applied to GPUs \cite{Shi2012}. Zirwes et al. (2018)~\cite{Zirwes_HPC_2018} proposed a conversion of reaction mechanisms into source codes to restructure the data of the kinetic mechanisms for efficient computation enabling compiler optimizations. Despite these strategies have resulted in noticeable computational cost reductions, ensuring a load balance strategy for general applications in premixed and non-premixed combustion will be of high value. All the previous strategies are based on a re-distribution of the load at system level, where the reaction rate evaluation and chemistry integration is redistributed according to the current load and an estimation of the stiffness using Message Passing Interface (MPI) standard. However, one of the fundamental issues associated to those strategies is the evaluation of the chemical stiffness, which is difficult to predict and can result in load imbalance. Moreover, when applying these methods it is necessary to synchronize the different processes and exchange data through MPI communication.

This paper is devoted to implement and analyze new load balancing mechanisms for chemistry integration. In particular, we propose utilizing the Dynamic Load Balancing (DLB) library \footnote{https://pm.bsc.es/dlb}~\cite{dlb,dlb2}, which allows reusing CPU-cores associated with idle MPI processes by other processes running on the same node. It is a load balancing mechanism based on transferring idle resources at the node level rather than transferring workload subsets through message passing. DLB acts as an automatic runtime mechanism transparent to the user and requires minimum changes in the source code (two lines in the present study). In fact, DLB can be combined with the workload transferring strategies mentioned above. DLB has already been successfully applied to increase the load balance for the assembly of the right-hand side terms in the Navier-Stokes equations~\cite{dlb_nastin}, the particle transport~\cite{dlb_particles}, coupled codes~\cite{dlb_coupled} and has been also used in different architectures~\cite{dlb_architectures}. Here, it is extended to optimize the chemistry part in reacting flow simulations.
The proposed solution with DLB, does not need to add extra data movement, because everything is done through the shared memory of the node. As it is a dynamic mechanism that reacts to the load imbalance, it does not need to predict the stiffness nor the computation load associated. And last but not least, it does not require a heavy implementation effort in the application.

The remaining of the paper is organized as follows. Section 2 describes the computational framework that is used for conducting the numerical simulations including the modelling and numerical descriptions used from the code Alya~\cite{Vazquez_JCS_2016}. Section 3 describes the computational environment in which these simulations are conducted and Section 4 details the assessment of this dynamic load balance strategy on representative problems in combustion science. Finally, the conclusions and directions of future work are given in Section 5.

\section{Modelling framework}
\label{}

\subsection{Governing transport equations}
\label{}

The simulation of reacting flows includes governing equations for chemical species along with energy, momentum and continuity. A low Mach number approximation of the Navier-Stokes is considered in this study for which the conservation of continuity and momentum read:

\begin{align}
    \label{eq:NS_cont}
    \frac{\partial \rho}{\partial t} + \boldsymbol{\nabla} \cdot \left( {\rho} \boldsymbol{u} \right) &=0,\\
    \label{eq:NS_mom}
    \frac{\partial (\rho \boldsymbol{u})}{\partial t} + \boldsymbol{\nabla} \cdot \left( \rho \boldsymbol{u} \boldsymbol{u} \right) &=
    -\boldsymbol{\nabla} p +\boldsymbol{\nabla}  \cdot \left( \mu \boldsymbol{\nabla}  \boldsymbol{u} \right),
\end{align}
where standard notation is used for all the quantities and $\rho$, $\boldsymbol{u}$, $p$ and $\mu$ represent the density, velocity vector, pressure and dynamic viscosity.

Regarding the evolution of the multicomponent gas, it can be expressed in terms of the transport equations for the individual species $Y_k$ given by:

\begin{equation}
    \label{eq:Yk}
    \frac{\partial (\rho Y_k)}{\partial t} + \boldsymbol{\nabla} \cdot \left( \rho \boldsymbol{u} Y_k \right) =
    \boldsymbol{\nabla}  \cdot \left( \rho D \boldsymbol{\nabla}  Y_k \right) +  \dot\omega_{Y_k} \quad \quad \quad k =1, \ldots, N_s.
\end{equation}
In this equation, $D$ is the diffusion mass coefficient, for which a unity Lewis assumption has been adopted, while $\dot\omega_{Y_k}$ denotes the chemical source term for species $Y_k$. $N_s$ is the number of species considered in the chemical mechanism.
Finally, the total enthalpy $h$ equation, in which heating due to viscous forces is neglected, reads as: 

\begin{align}
    \label{eq:h}
    \frac{\partial (\rho h)}{\partial t} + \boldsymbol{\nabla} \cdot \left( \rho \boldsymbol{u} h \right) &= \boldsymbol{\nabla}  \cdot \left( \rho  D \boldsymbol{\nabla}  h \right).
\end{align}

\subsection{Chemical integration}
\label{}

The chemical integration is one of the most computationally demanding parts in the integration of the governing equations due to the high non-linearity of the Arrhenius-type reaction kinetics. It is, therefore, clear that the integration method for chemistry may play an important role in the total time for the simulation, especially when detailed chemistry models are considered.

In this work, to reduce the stiffness of the integration of the species governing equations \ref{eq:Yk}, a splitting algorithm is used to separate the transport from the chemistry \cite{Ren2008}. The solution of the chemistry problem is achieved by the integration of the open source Cantera~\cite{Cantera2021} software as an external library in the multiphysics code Alya~\cite{Vazquez_JCS_2016}. A Fortran to C++ wrapper was created to integrate Cantera into Fortran for Alya, so internal functions from Cantera could be used in runtime. 
The reaction rates are obtained from in-built internal functions from Cantera, and the chemical integration is obtained using the CVODE algorithm \cite{Cohen1996}. A listing of the integration loop of the code is given in Listing~\ref{cod:alya}.

CVODE is a package written in C to solve IVPs (Initial Value Problems) defined by stiff and non-stiff ODEs in the form:

\begin{align}
    \label{eq:IVP}
    \boldsymbol{\dot{y}} = \boldsymbol{f}(t,\boldsymbol{y}),
\end{align}

with the initial conditions given by $\boldsymbol{y}(t=t_0) = \boldsymbol{y}_0$. 
In particular, the equations for chemistry integration are similar to those of system \ref{eq:IVP} but without the dependence on time:

\begin{equation}
\frac{d\boldsymbol{Y}}{dt} = \boldsymbol{f}(T, \boldsymbol{Y}),
\label{chem_eq}
\end{equation}
where $T$ is temperature, $\boldsymbol{Y}= (Y_1, \ldots, Y_N)^T$ is the vector of mass fraction for species. The initial conditions correspond to $T(t_0)=T_0$ and $\boldsymbol{Y}(t_0) = \boldsymbol{Y}_0$.

CVODE is in turn based on the ODE solver packages VODE and VODPK \cite{Brown1989} and solves previous IVP by the application of an implicit temporal scheme based on either Adams-Moulton formula or backward-differencing formula (BDF) methods, with the subsequent resolution of the non-linear equation by Newton's method. Depending on the form of the Jacobian matrix, dedicated functions for dense or banded matrices can be used allowing, moreover, the preconditioning of the linear system. 

The code used in Alya for chemical integration loop with no parallelization is gathered in Listing~\ref{cod:alya}.

\begin{lstlisting}[language=Fortran, caption=Chemical integration loop in Alya code., captionpos=b, commentstyle=\texttt{}, frame=lines, label=cod:alya]
do ipoin=1,npoin  
    if(reaction) call cvode_integration
end do
...
call MPI_Allreduce(...)
\end{lstlisting}

\subsection{Computational platform: Alya} 
\label{alya}
All the computational strategies presented in this paper have been implemented in Alya, the high performance computational multi-physics code developed at the Barcelona Supercomputing Center. Alya is developed using a
modular architecture that includes a module coupling to tackle complex multi-physics problems such as combustion simulations. Alya is written in Fortran and is designed for massively parallel  supercomputers; particularly it is 
 one of the twelve simulation codes of the Unified European Applications Benchmark Suite (UEABS)~\cite{ueabs}, being regularly tested on the European Tier-0 supercomputers.

The parallelization implemented in Alya comprises three levels: distributed memory for inter- and intra- node parallelism, shared memory for intra-node parallelism, and SIMD (Single Instruction Multiple Data) and SIMT (Single Instruction Multiple Thread) for CPU vectorization and GPU computing, respectively. The implementation of such strategy combines various programming models: MPI for inter-process message passing and synchronization, directives-based approaches (OpenMP, OmpSs, and OpenACC) for loops and task-based parallelism and GPU offloading, as well as CUDA for low-level optimized implementation of specific kernels.

The primary option for mesh partitioning is an in-house SFC-based partitioner~\cite{BORRELL2018}. Online redistribution is also performed to adjust the partition to run-time measurements. This last option has been exploited for co-execution on heterogeneous systems, where the partition is adjusted for a balanced execution using CPU and GPU devices simultaneously~\cite{BORRELL2020}. For task-based shared-memory parallelisms, a second-level decomposition is performed in which the size of the resulting subsets is of the order of $10^2$ elements. Finally, a data re-structuring is performed for vectorization: subsets of 8 to 32 elements are packed together to be executed in a SIMD model. The same strategy is used for GPU computing being, in this case, each pack, of the order of $10^5$ elements, launched to the GPU where the SIMT parallel model is exploited. 

\section{Description of the test cases}
\label{}

Two representative combustion problems are used here to evaluate the performance of DLB to ensure a load balancing strategy for the chemical integration.

\begin{figure}[htb!]
\centering
\begin{tabular}{c c}
  \includegraphics[width=0.8\linewidth]{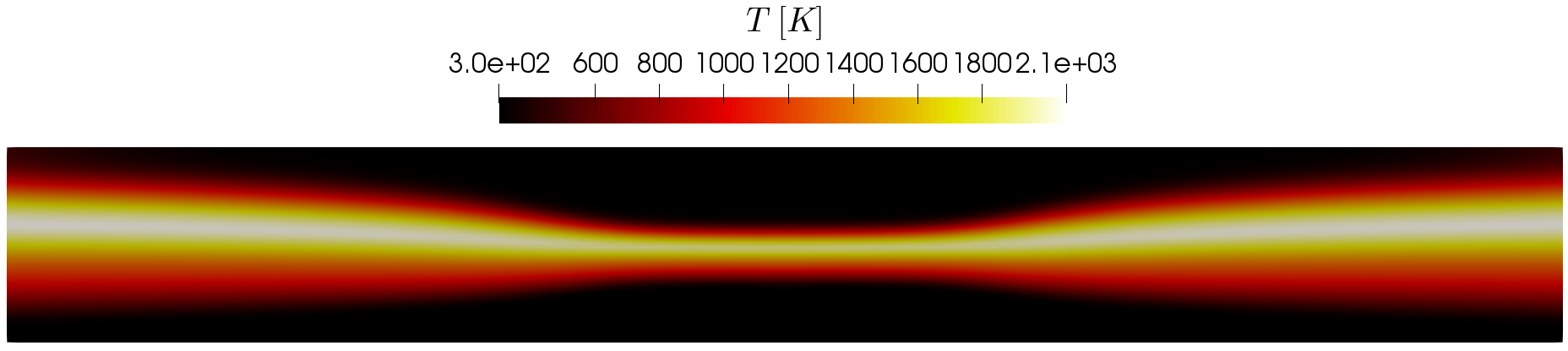} \\ \includegraphics[width=0.8\linewidth]{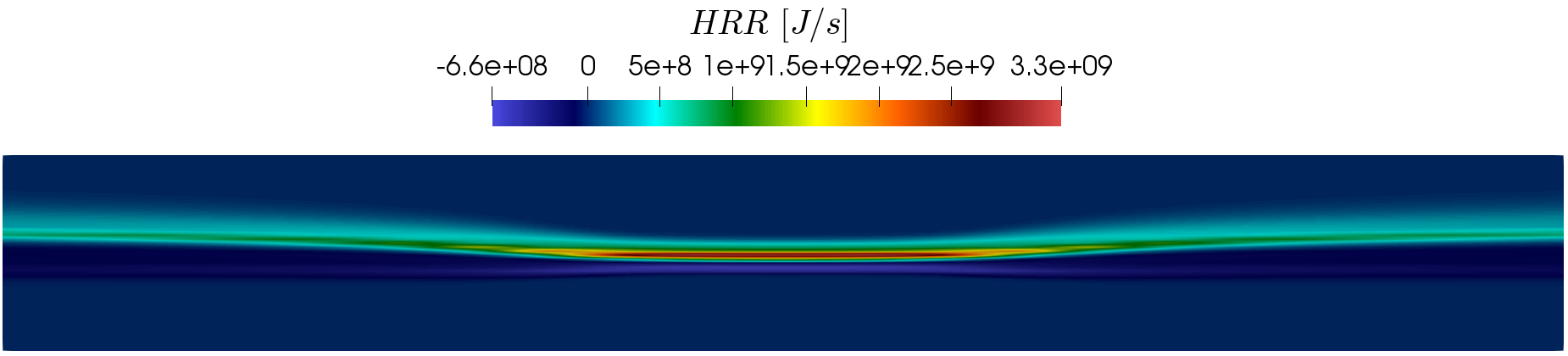}
\end{tabular}
  \captionof{figure}{Counterflow kerosene/air flame using the detailed reaction mechanism: temperature (top) and heat release rate (bottom).}
  \label{fig:flameCF}
\end{figure}

The first case corresponds to a laminar counterflow diffusion flame at atmospheric pressure and 298K air and fuel temperature. The flame in this configuration is characterized by a wide region where fuel and oxidizer are mixed and a reaction layer formed at the vicinity of the stoichiometric mixture fraction. Moreover, the flame is determined by the level of strain, which is given by the velocities of the two streams and the distance between the nozzles. A representation of the flame is shown in Fig.~\ref{fig:flameCF}. A two-dimensional domain with two different mesh sizes was used to evaluate the load balancing strategy using a single computing node (coarse mesh) and a multi-node calculation (fine mesh). In addition, to analyse the effect of load imbalance due to chemical integration, two reaction mechanisms with very different sizes were chosen to assess the performance of DLB when using detailed and reduced chemistry. The first reaction mechanism is a detailed chemical scheme for kerosene comprising 189 species and 1327 reactions~\cite{Kathrotia_GT2018-76997}, while the second is a 2-step reduced mechanism containing 6 species~\cite{FRANZELLI20101364}. 

\begin{figure}[htb!]
\centering
\begin{tabular}{c c}
  \includegraphics[width=0.7\linewidth]{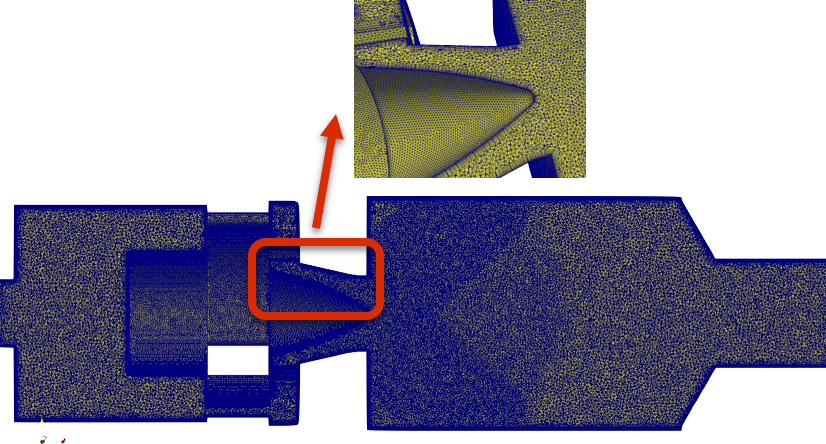} \\ \includegraphics[width=0.7\linewidth]{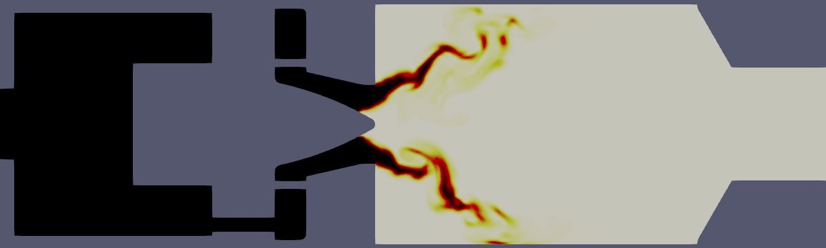}
\end{tabular}
  \captionof{figure}{Swirl-stabilized turbulent premixed flame using the detailed reaction mechanism: mesh resolution (top) and temperature (bottom).}
  \label{fig:flameSB}
\end{figure}

The second test case is a more realistic problem and features a turbulent premixed flame in a swirl-stabilized burner also known in the literature as the PRECCINSTA burner~\cite{Govert_FTaC_2018,Both_CAF_2020}. The operating point at equivalence ratio $\phi=0.67$~\cite{Both_CAF_2020} is taken as reference but changing the fuel to kerosene to allow for this comparison. For this case, the governing equations are filtered in space and the computational framework is adapted to run large-eddy simulations (LES). Details of the modelling and numerical approach are given in previous work~\cite{Both_CAF_2020} and are omitted here for brevity. For this test case, a finite rate model considering the filtered equations for LES are employed without accounting for Turbulence-Chemistry Interactions (TCI). This approach has been chosen due to its simplicity and with the aim to evaluate the load imbalance in LES and DNS applications. Notwithstanding, the conclusions drawn in the following related to the improvements of hybridization and DLB are expected to be valid for other turbulent combustion models that do chemical integration \emph{in situ} such as the Conditional Moment Closure (CMC)~\cite{KLIMENKO1999595}, Eddy Dissipation Concept~\cite{EDC_2020} or the Transported Probability Density Functions (TPDFs) models~\cite{Haworth2011}, greatly expanding the potential of DLB for advanced combustion simulations.
The computational domain includes the plenum, swirler and combustion chamber, and is composed by a hybrid mesh including prisms, tetrahedrons and pyramids. This is shown in Fig.~\ref{fig:flameSB} along with a sample snapshot of the temperature field. The same previous reaction mechanisms are also tested in this configuration. A summary of the computational cases and details of the mesh size and resolution is given in Table~\ref{tab:cases}.

\begin{table}[htb!]
    \centering
    \begin{tabular}{cccccc}
         Identifier &  Configuration    & Mechanism & Species & Reactions & Mesh (cells)  \\ \hline
         CF1  &  Counterflow  & Detailed & $189$ & $1327$ & 6k  \\
         CF2  &  Counterflow  & Reduced &  $6$ & $2$  & 6k  \\
         CF3  &  Counterflow  & Detailed & $189$ & $1327$ & 52k  \\
         CF4  &  Counterflow  & Reduced & $6$ & $2$  & 52k  \\
         SB1  &  Swirl burner & Detailed & $189$ & $1327$ & 24M  \\
         SB2  &  Swirl burner & Reduced & $6$ & $2$ & 24M 
    \end{tabular}
    \caption{Description of the computational cases.}
    \label{tab:cases}
\end{table}

For both of the cases, two metrics were obtained, namely, the time elapsed for the chemical integration in a Runge-Kutta sub-step and the total time required for the integration of the whole time step. The analysis includes single-node and multi-node tests on the counterflow configuration aiming to evaluate the effect of the hybridization, grain size and the optimization with DLB. After the correct identification of the optimal DLB parameters for the counterflow flame, the same settings are applied to the turbulent swirl burner case in order to evaluate this methodology on production runs. Details of the analysis are given in the next subsections.

\section{Computational background}
\label{sec:comp_backg}

\subsection{Performance analysis}
\label{sec:perf}

We start the study with a performance analysis of the execution which is carried out following the POP\footnote{https://pop-coe.eu/} methodology~\cite{pop,pdp}. We use the BSC performance analysis tools\footnote{https://tools.bsc.es}: EXTRAE~\cite{extrae} to generate execution traces and Paraver~\cite{paraver} to visualize them.

\begin{figure}[htb!]
\centering
  \includegraphics[width=0.7\linewidth]{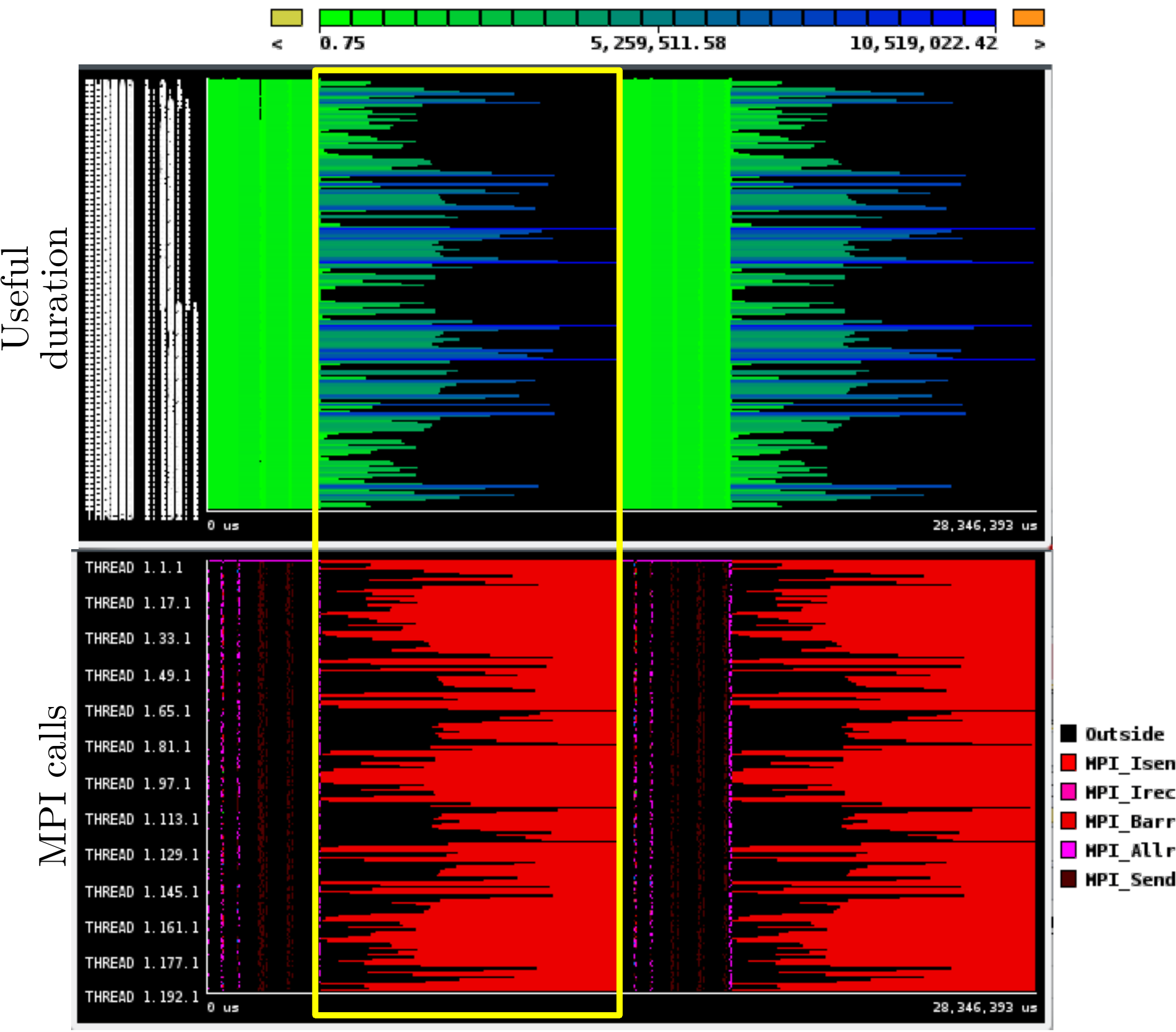}
  \caption{Paraver trace showing two time steps for the detailed chemistry simulation of the counterflow flame CF1.}
  \label{fig:trace_detailed}
\end{figure}

In Figure~\ref{fig:trace_detailed}, we can see two timelines from Paraver. Paraver timelines show in the x axis the time and each row corresponds to one MPI process. The MPI processes are ordered going from MPI rank 0 in the top of the view to the highest rank in the bottom of the view. The color can identify different metrics depending on the view that is selected.

On the top trace, we can see in a color bar the duration of the useful computation. We consider useful computation when the process or thread is doing computation and not waiting in an MPI call. On the bottom trace, we show the MPI call, in this case black color corresponds to useful computation, because it is outside an MPI call. Both traces are depicted using the same timescale.

For illustrative purposes, in this view we show two time steps for one of the cases simulated in this work, namely, the detailed chemistry simulation for a counterflow flame. In this case the execution is done using 192 MPI ranks. We have marked with a yellow square one of the chemical integration loops.

We can observe the important load imbalance present in this section of the execution, where some MPI ranks have a higher load than other ones. This produces an important time spent waiting in the MPI\_Barrier call by some MPI processes.

\begin{figure}[htb!]
\centering
  \includegraphics[width=0.7\linewidth]{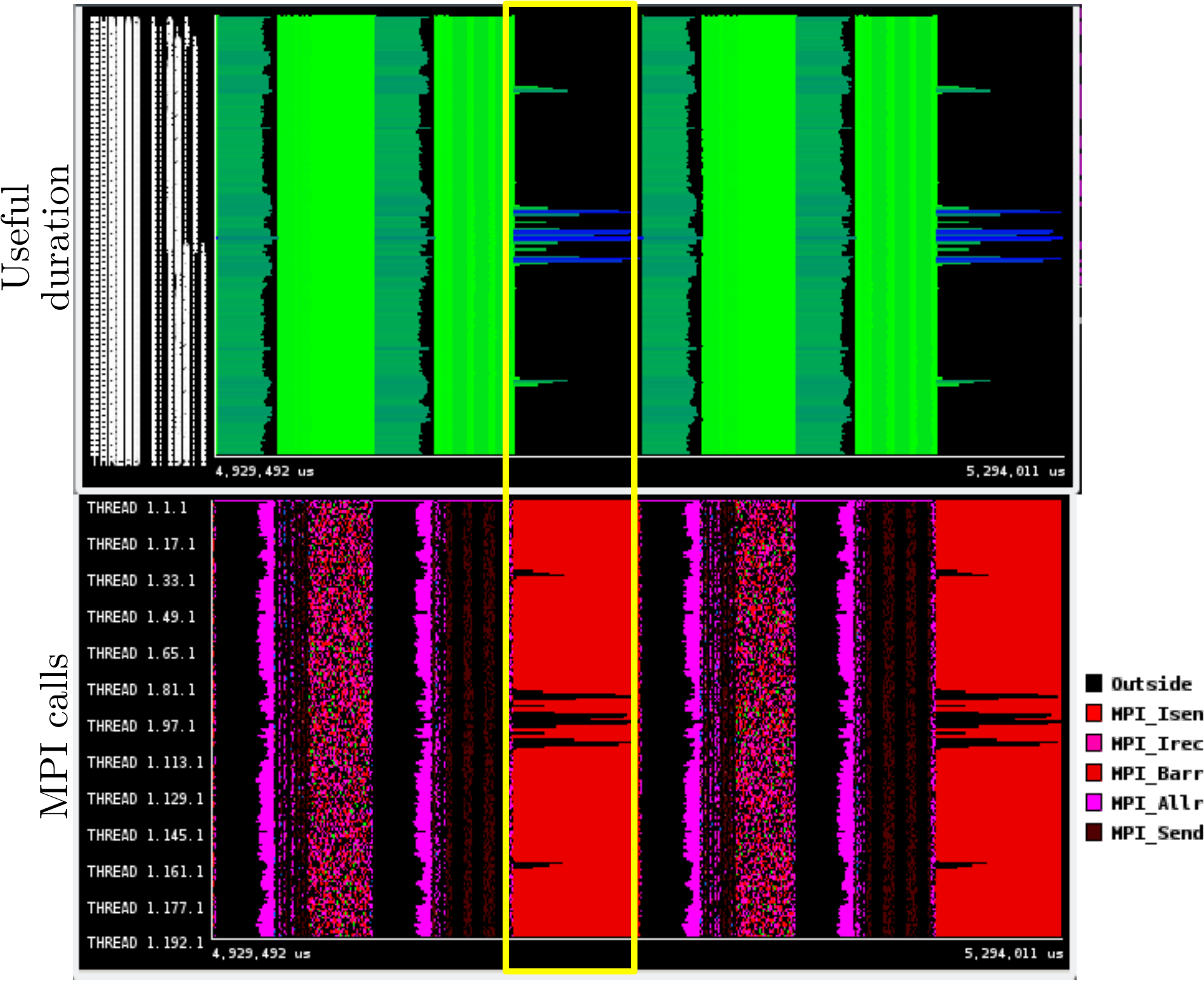}
  \caption{Paraver trace showing two time steps for the reduced chemistry simulation of the counterflow flame CF2.}
  \label{fig:trace_reduced}
\end{figure}

In Figure~\ref{fig:trace_reduced} we see the same analysis for the simulation of a counterflow flame with reduced chemistry. We can see the 192 MPI ranks in the different rows and two time steps in the time scale (x axis). The top trace shows the duration of the useful computation, while the bottom one shows the MPI calls being executed. 

This use case also presents a high load imbalance in the chemical integration phases being one of the integration loops again marked with a yellow square. We can observe that compared to the detailed use case, the relative duration of the chemical integration respect to the duration of the whole time step is lower in the case of the reduced chemistry. Moreover, we can observe that the load imbalance pattern is quite different, in the reduced chemistry the most loaded processes are very localized in a few MPI processes.

In order to quantify the efficiency loss due to the load imbalance we use the POP metrics. In Table~\ref{tab:efficiencies} we show the parallel efficiency metrics obtained by the two use cases analyzed, CF1 and CF2. We compute the efficiency metrics based on the whole timestep and on the chemical integration loop.

\begin{table}[htb!]
\centering
  \includegraphics[width=0.9\linewidth]{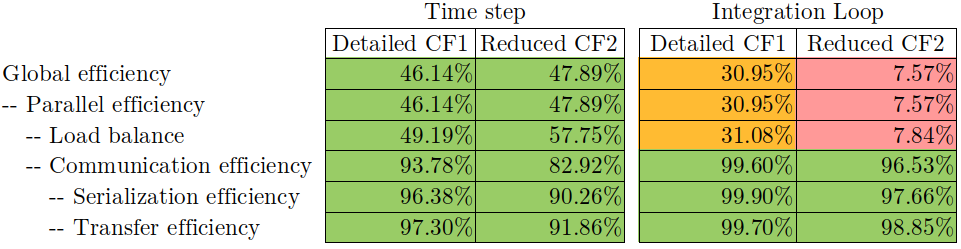}
  \caption{POP efficiency metrics obtained for the detailed (CF1) and reduced (CF2) chemistry simulations.}
  \label{tab:efficiencies}
\end{table}

When taking into account the whole time step both inputs show a similar Load Balance, 49.19$\%$ and 57.75$\%$ for CF1 and CF2 cases, respectively.
If the CF1 case is considered, this means that from all the computing power used for this simulation only $49\%$ is used to do useful work, the remaining $51\%$ is lost due to load imbalance. When we consider only the chemical integration loop we see a big difference between the two physics, the detailed one (CF1) presents a Load Balance of $31\%$ while the reduced one (CF2) shows a Load Balance of $7\%$.

Based on the performance analysis, we see that the main factor limiting the scalability for this simulation is the load imbalance. The nature of the problem suggests that the best approach is to use a dynamic mechanism because we see that when changing the complexity of the chemical reactions the load balance pattern changes. In order to address the dynamically the different imbalances our proposal is to use the DLB library.

\subsection{DLB Library}
\label{sec:dlb}

The Dynamic Load Balancing Library (DLB) is a framework that aims at improving the load balancing of hybrid applications. It is transparent to the application and it is integrated with MPI, OpenMP and OmpSs~\cite{ompss}.

The load balancing algorithm used in the present work is LeWI (Lend When Idle). The main idea of this algorithm is to use the computational resources (cores) that are idle when an MPI process is waiting in a blocking MPI call to speed up another process in the same node. This can be done taking advantage of the malleability of shared memory programming models such as OpenMP or OmpSs.

\begin{figure}[htb!]
\centering
  \includegraphics[width=0.7\linewidth]{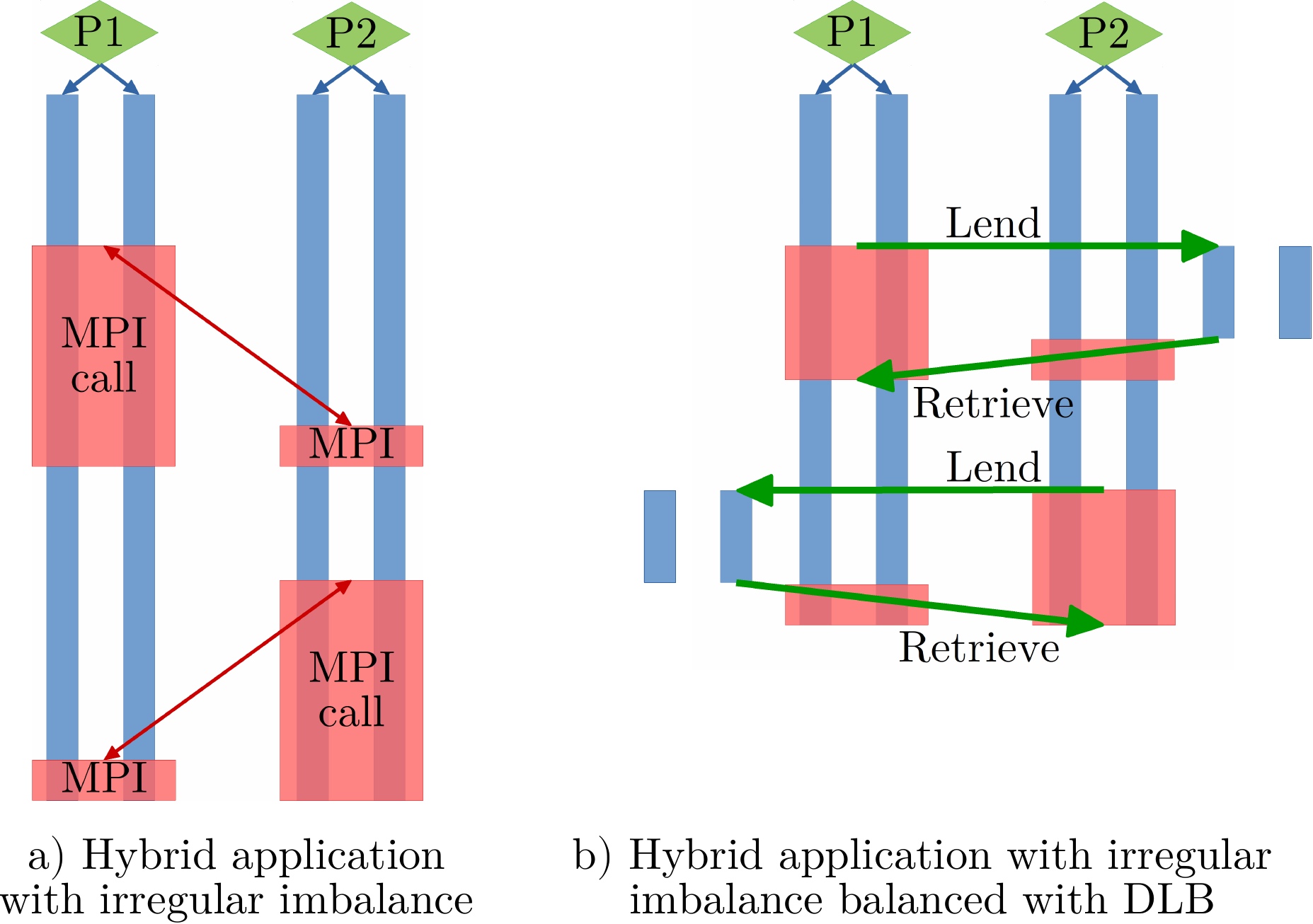}
  \caption{Example of hybrid application load balanced with DLB.}
  \label{fig:dlb}
\end{figure}

In Figure~\ref{fig:dlb} we show an example of a hybrid application load balanced using DLB and LeWI. On the left hand side of Figure~\ref{fig:dlb} , we can see an application with two MPI processes and two OpenMP threads each one. The application presents a load imbalance problem, MPI process P1 finishes its computation faster and waits in an MPI blocking call for MPI process P2. Moreover, the load balance is irregular, because on the next iteration P2 finishes its computation faster and waits for P1 to finish its computation.

On the right hand side of the Figure~\ref{fig:dlb} we can observe how DLB can speed up this application. When MPI process P1 reaches the MPI blocking call, it lends its computational resources to MPI process P2. At this point P2 is able to use 4 OpenMP threads, finishing the remaining work faster. When P1 reaches the end of the blocking MPI call it retrieves its original resources, thus continuing the execution with the original 2 OpenMP threads. In the following iteration, when process P2 reaches the blocking MPI call, it lends its resources to P1 and now P1 is able to use 4 OpenMP threads.

\subsection{Implementation}
\label{sec:impl}

In order to use DLB we need a second level of parallelism based on a shared memory programming model. We can add this second level of parallelism only in the regions exhibiting a load imbalance, thus avoiding the need for parallelizing the whole application. With this approach the second level of parallelism will be only used to balance MPI processes load and, during the remaining of the execution, each MPI process will run just with one thread.

As the second level of parallelism is added only to be used with DLB we decided to use OmpSs because the current version of DLB offers a better integration with OmpSs. Therefore, the first step is to add the necessary pragmas in the code to allow the parallelization with OmpSs. For the particular case of chemical integration (see Listing~\ref{cod:alya}), Listing~\ref{cod:hybrid} gathers the required code for its parallelization with OmpSs.

\begin{lstlisting}[language=Fortran, caption=Chemical integration loop parallelized with OmpSs., captionpos=b, commentstyle=\texttt{}, frame=lines, label=cod:hybrid]
!$omp taskloop private(ipoin) default(shared) &
!$omp& grainsize(OMPSS_GRAINSIZE)
do ipoin=1,npoin  
    if(reaction) call cvode_integration
end do
...
call MPI_Allreduce(...)
\end{lstlisting}

One OmpSs pragma is added (\texttt{taskloop} in lines 1 and 2) to distribute the iterations of the loop over the points (line 3) into tasks that can be executed by the different threads. The necessary data sharing clauses are added (\texttt{private} and \texttt{shared}) for the correct execution of the code. Additionally the clause \texttt{grainsize} is used to tell the runtime how many points will be computed in each parallel task. At this point the grain size variable points to a global variable that can be modified for the different runs. In section~\ref{sec:eval_hybrid} the impact of the grain size in the performance of the hybrid parallelization will be studied.

The next step is to enable DLB load balancing for the execution. This can be done by \textit{preloading} the DLB library and setting the necessary environment variables for the OmpSs runtime and DLB. In Listing~\ref{cod:script} we show an example of the lines that can be added to the submission script to enable DLB.

\begin{lstlisting}[language=bash, caption=Example of script to use DLB., captionpos=b, commentstyle=\texttt{}, frame=lines, label=cod:script]
export DLB_ARGS="--lewi"
export DLB_HOME=${path_to_dlb_instalation}$
export LD_PRELOAD=${DLB_HOME}/lib/libdlb_mpi.so
export NX_ARGS="--enable-dlb --enable-blocking"
#Run application as usual
\end{lstlisting}

However, in this case, as we wanted to enable DLB in a very delimited part of the code we decided to use the API (Application Programming Interface) offered by DLB. For this purpose, we need to add a single call to DLB before the collective communication that was suffering from the load imbalance. Therefore, the DLB call that we add is \texttt{DLB\_Barrier} that acts as a barrier only between the MPI processes in the same node and the processes that enter into the barrier, instead of doing a busy wait, lend their resources to DLB. With this approach there is no need to intercept all the MPI calls done by Alya reducing, in consequence, the overhead.

\begin{lstlisting}[language=Fortran, caption=Chemical integration loop with OmpSs and DLB., captionpos=b, commentstyle=\texttt{}, frame=lines, label=cod:dlb]
!$omp taskloop private(ipoin) default(shared) &
!$omp& grainsize(OMPSS_GRAINSIZE)
do ipoin=1,npoin  
    if(reaction) call cvode_integration
end do
...
call DLB Barrier( )
call MPI_Allreduce(...)
\end{lstlisting}

Listing~\ref{cod:dlb} shows the code of the chemical integration loop with the OmpSs parallelization and the call to the API to enable DLB (line 7). With these instructions the code can be run using DLB library.

\begin{figure}[htb!]
\centering
  \includegraphics[width=0.8\linewidth]{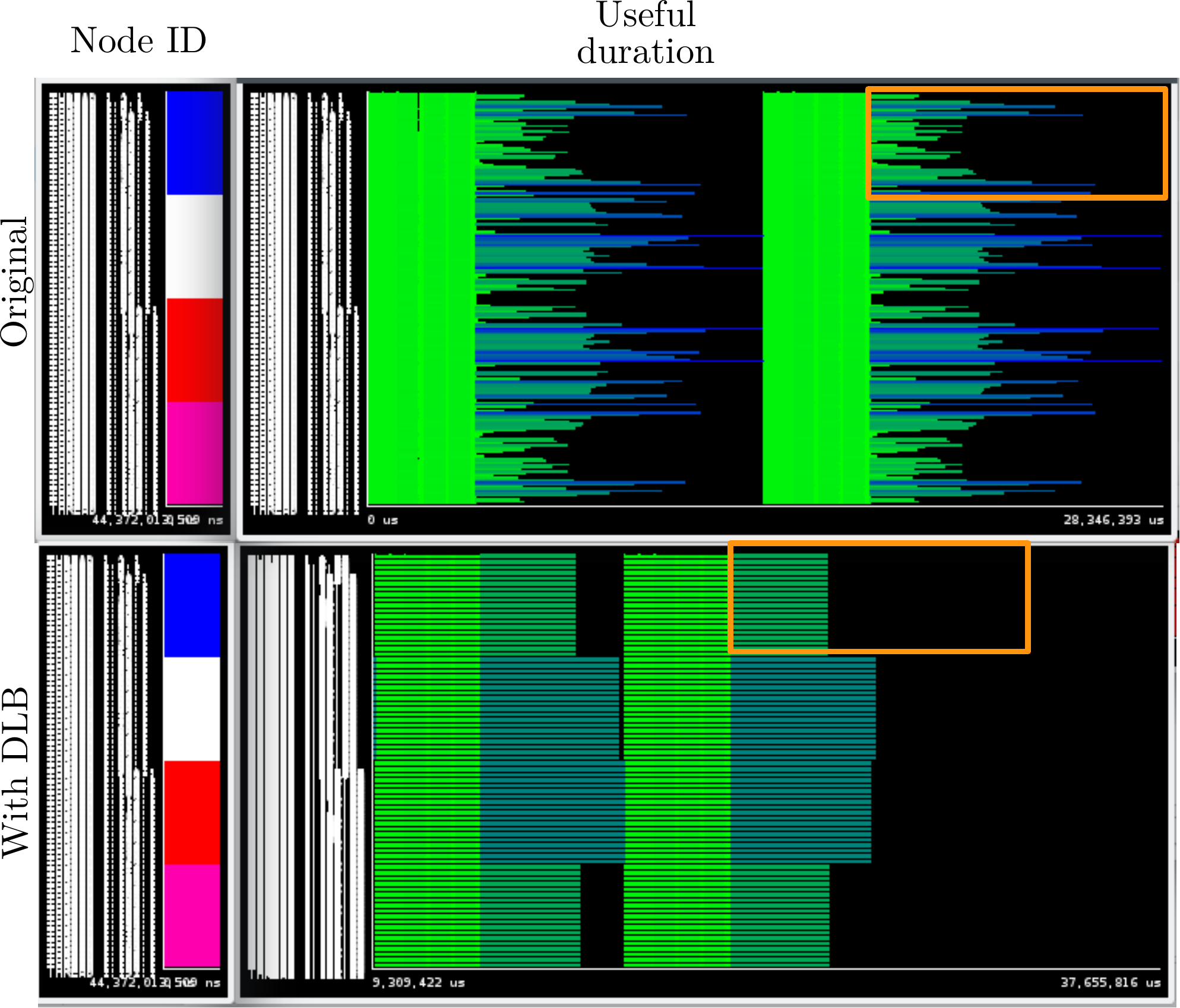}
  \caption{Paraver trace of two steps of the detailed chemistry showing the effect of DLB.}
  \label{fig:trace_dlb}
\end{figure}

In Figure~\ref{fig:trace_dlb} we show two traces to illustrate the use of DLB. Both traces are obtained from the detailed chemistry of the counterflow. The top trace is the original simulation using 192 MPI ranks (4 nodes from Marenostrum4) in pure MPI mode. The bottom trace is the same run using DLB and both traces are at the same time scale, so we can clearly see the benefit in execution time achieved by the use of DLB. We can also observe how DLB is able to load balance the chemical integration loop. In the left hand side we show in code color the node in which each MPI process is running and we can clearly observe how DLB balances load within each computing node. 

\begin{figure}[htbp!]
\centering
  \includegraphics[width=0.8\linewidth]{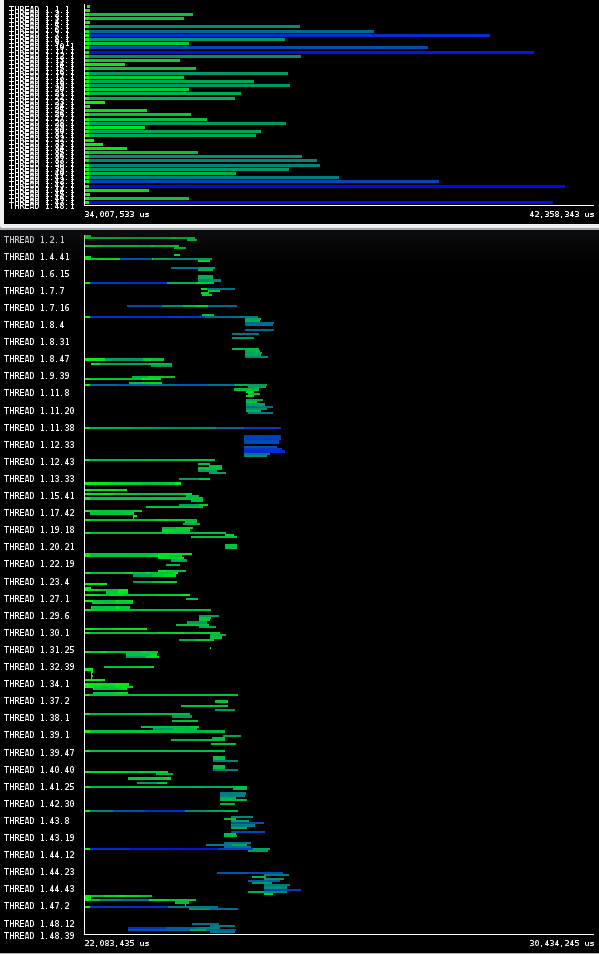}
  \caption{Zoom in a Paraver trace of two steps of the detailed chemistry showing the effect of DLB.}
  \label{fig:trace_dlb_zoom}
\end{figure}

In Figure~\ref{fig:trace_dlb_zoom} we see a zoom corresponding to the orange squares of Figure~\ref{fig:trace_dlb} to show in detail how DLB acts to load balance the simulation. It corresponds to the processes running in the first node and the region of the chemical integration. The top trace corresponds to the original execution and the bottom trace to the same run with DLB. In the top trace each line corresponds to one MPI process. In the bottom trace each line corresponds to one OmpSs thread, and they are grouped by MPI process, we can see how the number of active threads at one given point in time changes. If a vertical line is drawn at any point of the trace it will cross 48 active threads, because there are 48 cores available in the node.

When the less loaded processes finish their work they lend their resources and another process is able to spawn an extra OmpSs thread. This allows the execution with DLB to be more than $2\times$ faster than the original one for the region being analyzed.

\section{Evaluation}
\label{sec:eval}

In this section we present the performance evaluation of the code challenges shown in Section~\ref{sec:impl}. We organize the section in an incremental approach, we start analyzing the hybridization of the code and the impact of the grain size, then we show the performance benefit of applying DLB and also the impact of the grain size to the DLB performance. Based on the best configuration determined in those sections we then show the performance improvement when using multiple nodes. 

\subsection{Environment}
\label{sec:env}
The experiments carried out in this work have been performed on MareNostrum4~\cite{MN4}. This supercomputer is based on Intel Xeon Platinum processors from the Skylake generation. It is a Lenovo system composed of SD530 Compute Racks, an Intel Omni-Path high performance network interconnect and running SuSE Linux Enterprise Server as operating system. Compute nodes are equipped with 2 sockets Intel Xeon Platinum 8160 CPU with 24 cores each running at 2.10GHz for a total of 48 cores per node and 96 GB of main memory (1.88 GB/core).

We have used the Intel compiler 2017.4 and IMPI 2017.4 as MPI library. For all the experiments we use the master branch of Alya integrated with the Cantera library version 2.1. For the optimization using DLB we have used DLB almost 3.0 and OmpSs 19.06.

All the results gathered in this section are averages of 5 runs. As in all the cases, the standard deviation between the different runs is below 5\%, the error bars are not shown in the plots.

\subsection{Single-node testing}
\label{}

The main aim of this section is to determine the impact of the code hybridization with OmpSs on the performance as well as to be able to distinguish in a second stage the optimizations provided by DLB. Also, we study in detail the impact of the grain size when parallelizing with OmpSs to understand the importance of this factor on the computational cost and to try to find its optimal value or range of values.

The first set of tests comprises the analysis of the time reductions by the use of hybridization and DLB on the counterflow flame configurations CF1 and CF2.

\subsubsection{Hybridization and grain size study}
\label{sec:eval_hybrid}

In this section, we analyze the performance of the hybrid code, outlined in Listing~\ref{cod:hybrid}, versus the original MPI-only implementation. 

\begin{figure}[htbp!]
\centering
  \includegraphics[width=0.7\linewidth]{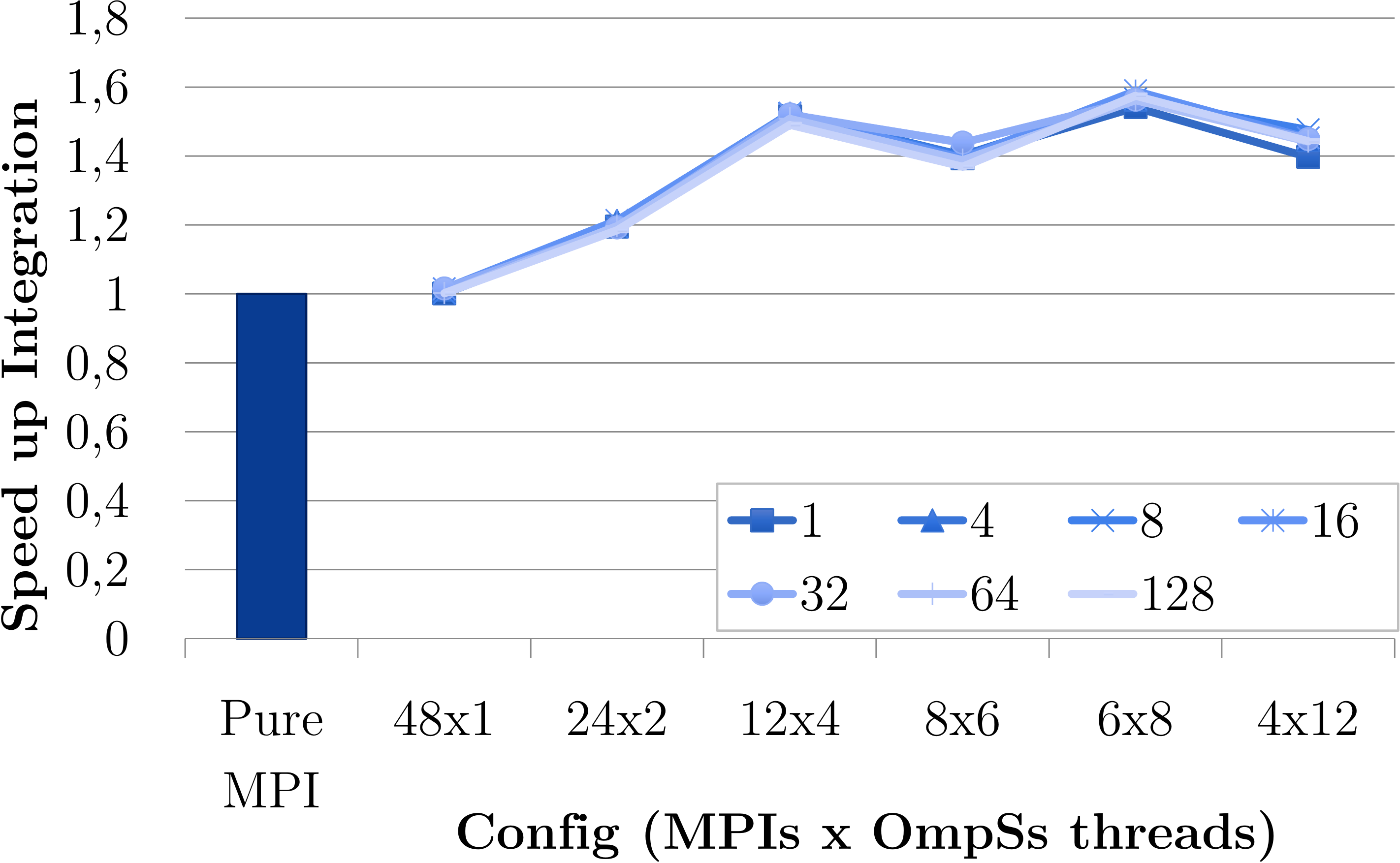}
  \captionof{figure}{Comparison of time for detailed chemistry integration (case CF1) between only MPI and hybridization with different number of threads for different grain size values.}
  \label{fig:1node_hybrid_detailed}
\end{figure}

Figure~\ref{fig:1node_hybrid_detailed} presents a numerical study for the counterflow flame considering the detailed chemistry CF1 case
using a single node of the MareNostrum IV supercomputer. In the X axis we can see different configurations of MPI processes and OmpSs threads, e.g. $12\times4$ corresponds to 12 MPI processes and 4 OmpSs threads each. Note that the product of these combinations is always 48 as all the results in the same plot are using the same number of computational resources (48 cores). In the Y axis we can see the speed up of the integration loop with respect to the MPI only version with 48 MPI processes (the pure MPI only is depicted as a bar in the plot as a reference). The different series represented with lines correspond to the hybrid version with different values for the grain size.

We observe that the MPI-only execution takes the same time as the $48\times 1$ configuration; therefore, the activation of OmpSs does not generate an appreciable overhead. Moreover, the trend is that by increasing the number of OmpSs threads per node, the chemistry integration cost is reduced. As there are no inter-process communications, the main reason for this acceleration is the implicit load-balancing obtained from the shared memory taskification: stiff and non-stiff tasks are assigned to  OmpSs threads as they are completed. 
Finally, we observe that results are almost independent of the granularity as the overlapping of the different lines shows. This means that there is no relevant overhead added by the hybridization of OmpSs and also that the granularity of tasks is small enough to provide parallelism and work for all the threads. 
Only the granularity of one element per task shows a very slight reduction of the speedup for the configuration $12\times 4$, this is because in this case 12 OmpSs threads are asking for work to the runtime system, as the tasks are very small we can start to see some congestion in the runtime when accessing the queue of work. 
All in all, $6\times 8 $ is the best configuration, and the speedup obtained versus the pure MPI code is $1.6\times$.

\begin{figure}[htbp!]
\centering
  \includegraphics[width=0.7\linewidth]{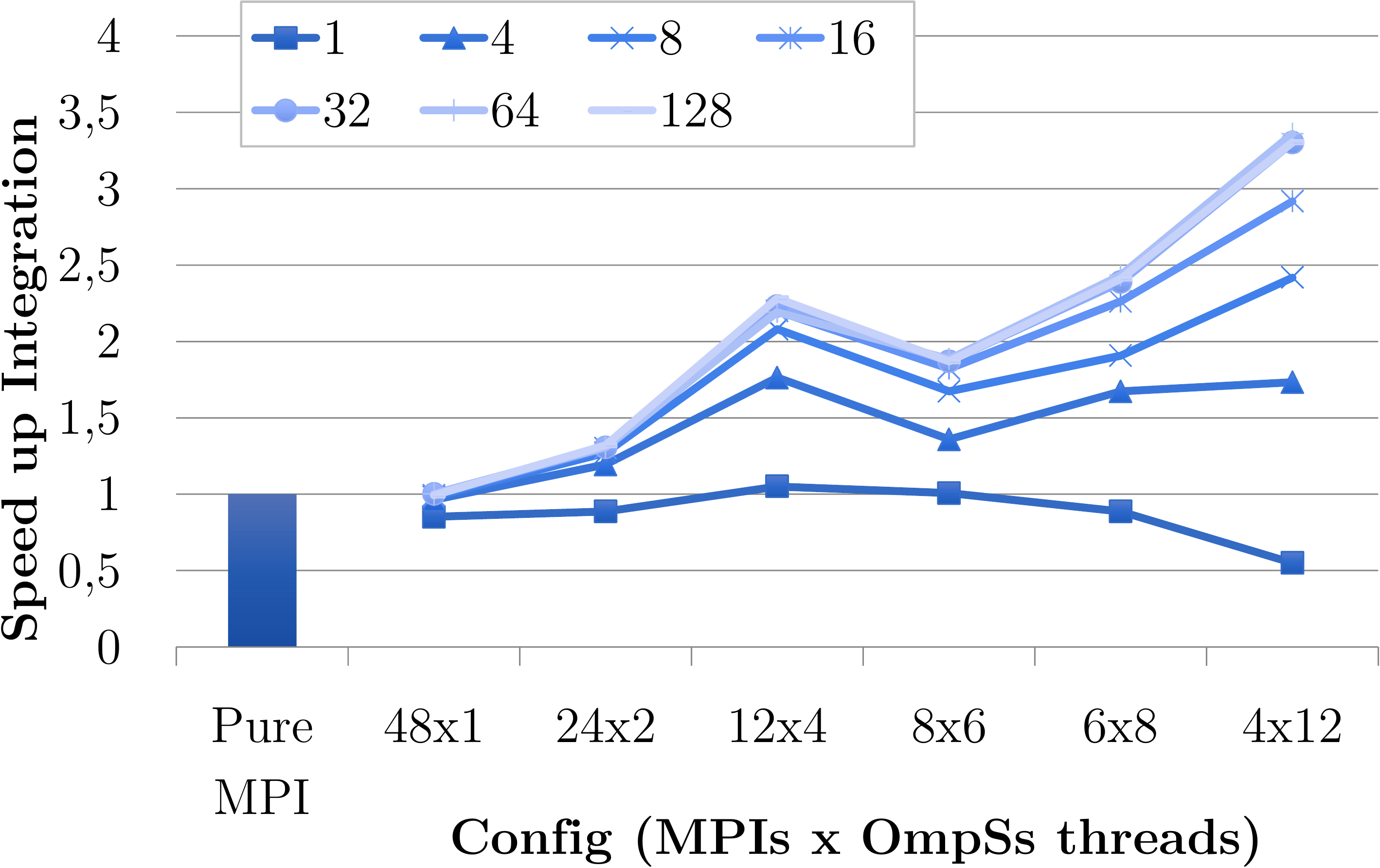}
  \captionof{figure}{Comparison of time for reduced chemistry integration (case CF2) between only MPI and hybridization with different number of threads for different grain size values.}
  \label{fig:1node_hybrid_reduced}
\end{figure}

In Figure~\ref{fig:1node_hybrid_reduced} we see the same results for the reduced chemistry CF2 case for which qualitative differences arise with regards to the detailed chemistry case CF1. In this case, a direct integration of the chemistry can be achieved with much fewer iterations in the CVODE solver, obtaining much lower overall computing times compared to the detailed chemistry case CF1. On the one hand, the grain size significantly affects the performance. In this case, unlike for the detailed chemistry, the cost of the integration of a single element is very low, so a couple of them need to be gathered per task to counterbalance the OmpSs overhead. On the other hand, we observe that the speedup rises up to $3.4\times$ and the best result is obtained with the configuration $12\times 4$. This higher speedup comes from an initially higher imbalance as we have seen in the performance analysis section (Section~\ref{sec:perf}),  whose correction produces more noticeable effects.  

As the problem is entirely governed by chemistry, the evaluation of the global rates is local and can strongly benefit from parallelization. This study proposes the use of hybrid code that uses the parallel programming model OmpSs to improve the load balance and reduce the computational cost. 
Moreover, the OmpSs parallelization does not add a significant overhead although, in cases with a very low computational load for the chemical integration, a small grain size can add overhead when the number of threads is increased. In these cases, the grain size might have an important impact on the performance and a mid to high value of the grain size can appreciably mitigate such overhead.

\subsubsection{DLB evaluation}
\label{}

In this section, we study the benefit of using DLB as an additional load balancing mechanism intra-node and how the grain size affects in these simulations. 

\begin{figure}[htbp!]
\centering
  \includegraphics[width=0.7\linewidth]{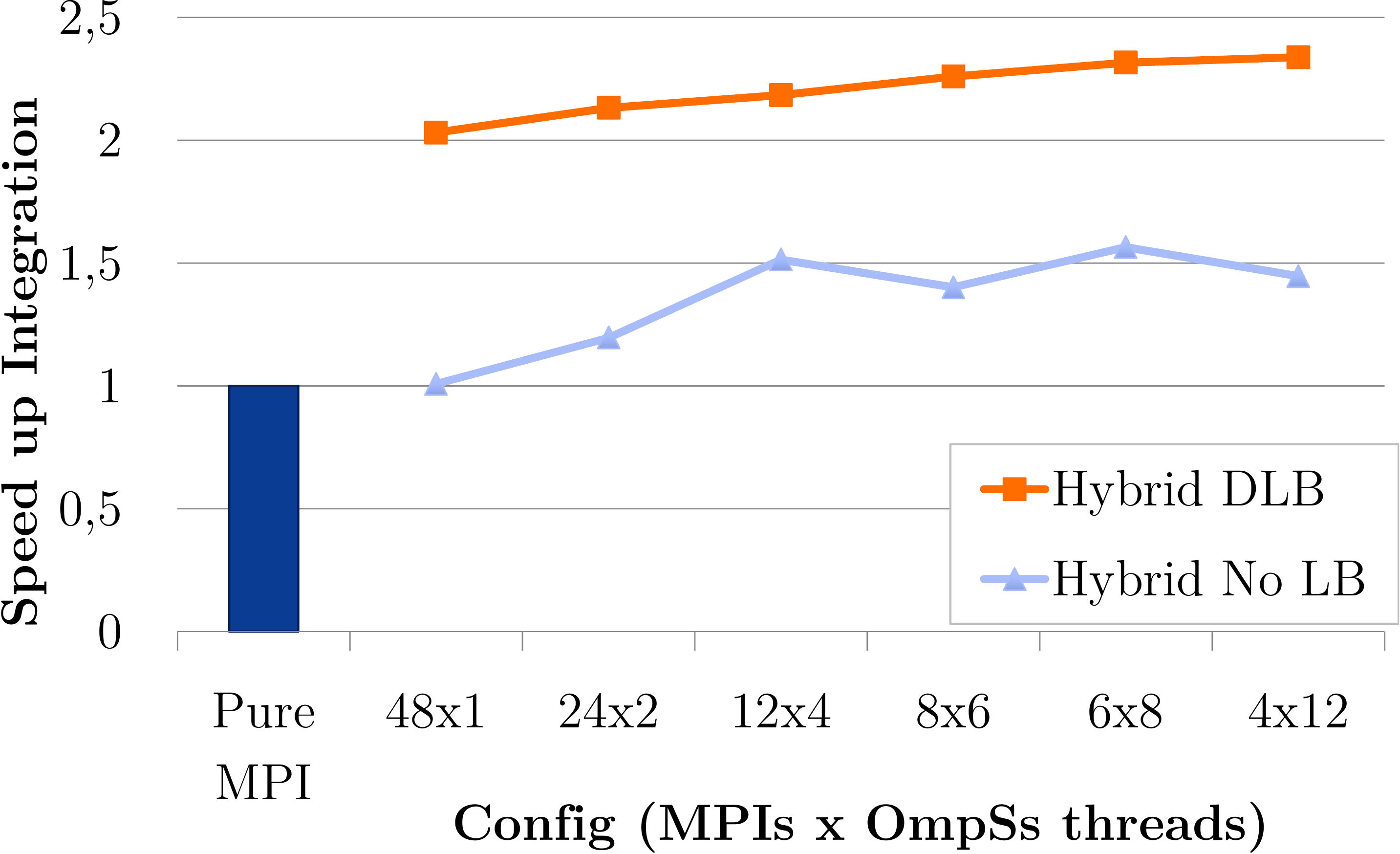}
  \captionof{figure}{Comparison of time for detailed chemistry integration (case CF1) between only MPI and hybridization with and without DLB for different number of threads with grain size 32.}
  \label{fig:1node_DLB_det}
\end{figure}

In Figure~\ref{fig:1node_DLB_det}, we show the speedup obtained by the hybrid version with and without DLB with respect to the MPI-only version (shown as a blue bar in the plot) when running in one node of Marenostrum4. In the X axis we can see the different configurations of MPI threads and OmpSs threads to fill one node of 48 cores. For all the hybrid executions we use a grain size of 32, because that is the minimal size that does not show any significant 
overhead in this problem. 

Adding DLB for the detailed chemistry integration as shown for CF1, Figure ~\ref{fig:1node_DLB_det}, generates a speedup of up to $2.3\times$  versus the MPI-only implementation, and up to $1.5\times$ versus the best hybrid configuration. This indicates that the use of DLB can improve the performance and address the imbalance further than only the hybridization of the code. It is relevant that with the configuration $48\times1$, we obtain results that are not far from optimal, a speed up of $2\times$. It means that we can get most of the performance without requiring an overall hybridization of the code; only one thread per MPI process needs to be activated in the zones where DLB is used. It is important to notice that the line corresponding to the executions with DLB (orange) is flatter than the one corresponding to the hybrid code. This means that the use of DLB makes the performance less dependent on the hybrid configuration, thus, it relieves the pressure from the user to decide the optimal configuration of MPI processes and threads.

\begin{figure}[htbp!]
\centering
  \includegraphics[width=0.7\linewidth]{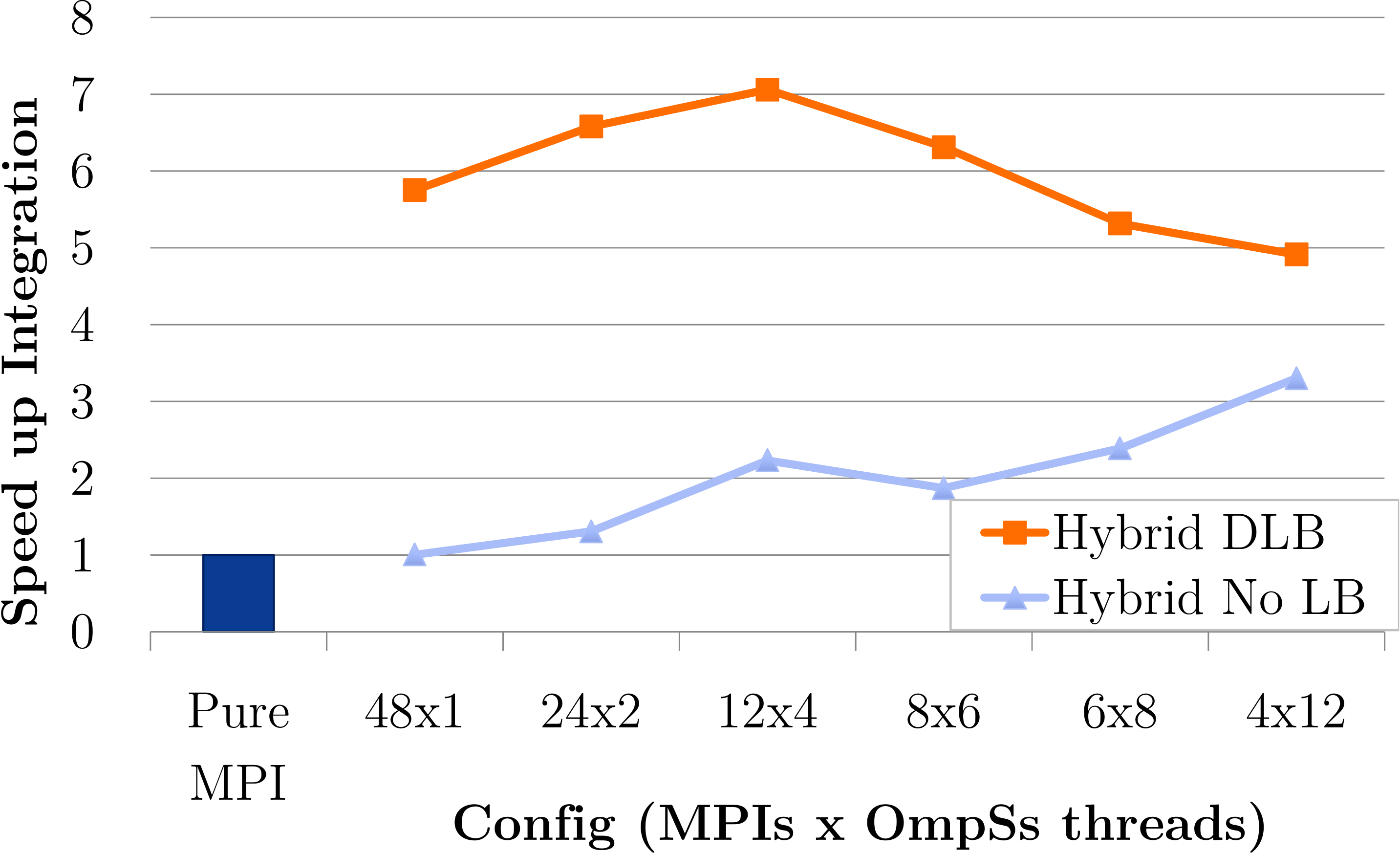}
  \captionof{figure}{Comparison of time for reduced chemistry integration (case CF2) between only MPI and hybridization with and without DLB for different number of threads with grain size 32. }
  \label{fig:1node_DLB_red}
\end{figure}

For the reduced chemistry CF2, we can see the results obtained with DLB in Figure~\ref{fig:1node_DLB_red}. As in the previous plot, we show the speedup over the MPI-only version (Y axis), when using different configurations of MPI processes and OmpSs threads (X axis). It is observed that the speedup 
versus the MPI-only version rises up to $7\times$, which represents an additional $2\times$ acceleration versus the best hybrid option. 

The speedup obtained by DLB in the reduced chemistry integration is higher than in the detailed one, as
the load imbalance is also higher in the reduced 
chemistry case. Note that the speedup that DLB can obtain is related to the existing load imbalance of the application.
As DLB re-distributes the tasks by the idle time from the processors, there is no need to predict the stiffness from the chemical problem as this is handled by DLB.

A second important aspect to consider is the impact of the grain size on the DLB performance in these applications.
\begin{figure}[htbp!]
\centering
\includegraphics[width=0.7\linewidth]{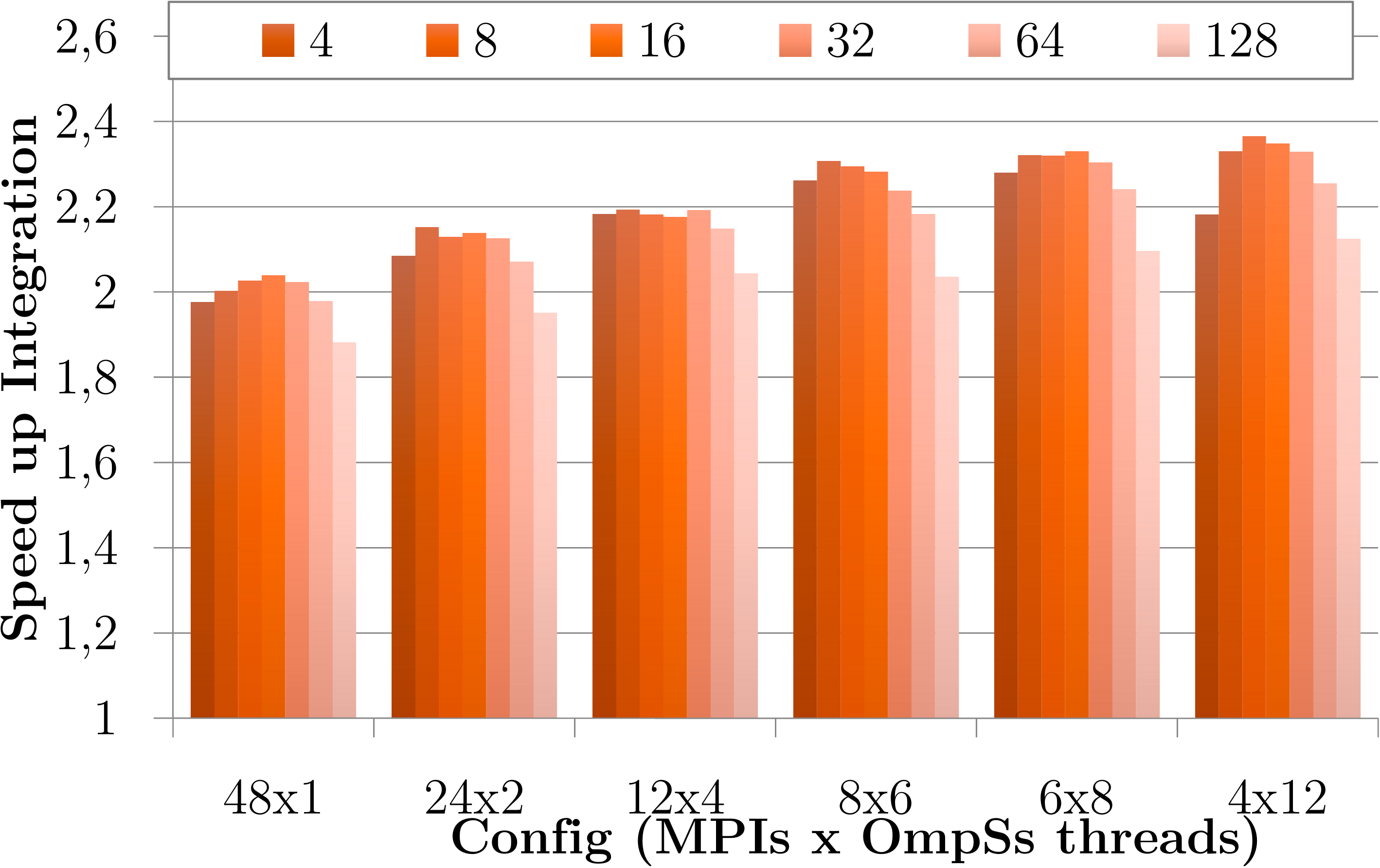}
  \captionof{figure}{Comparison of grain size impact when using DLB for detailed chemistry integration (case CF1) for different number of threads. }
  \label{fig:1node_DLB_grainsize_det}
\end{figure}
Figure~\ref{fig:1node_DLB_grainsize_det} shows the speedup versus the MPI-only version of the code (y axis) for the detailed chemistry case. We can see the different configurations of MPI processes and OmpSs threads to fill a node of 48 cores. The different series represent the different grain sizes used by the hybrid code and DLB. To understand the results, a trade-off between two aspects needs to be considered: i) the imbalance can be better reduced with thinner granularity; ii) the overhead of OmpSs is inversely proportional to the task size. We can see that the optimal grain size is always lower than or equal to 32. In the detailed case, the imbalance dominates the trade-off because the average cost of the chemistry integration per element is large compared to the OmpSs overheads. Considering the optimal grain size for each configuration, the speedup obtained ranges between $2\times$ and $2.3 \times$  with the $48\times 1$ configuration, we achieve $86\%$ of the maximum speedup. Moreover, a low variance 
between different configurations is observed, which reinforces previous observations about the use of DLB to isolate the performance from the selected configuration.

\begin{figure}[htbp!]
\centering
\includegraphics[width=0.7\linewidth]{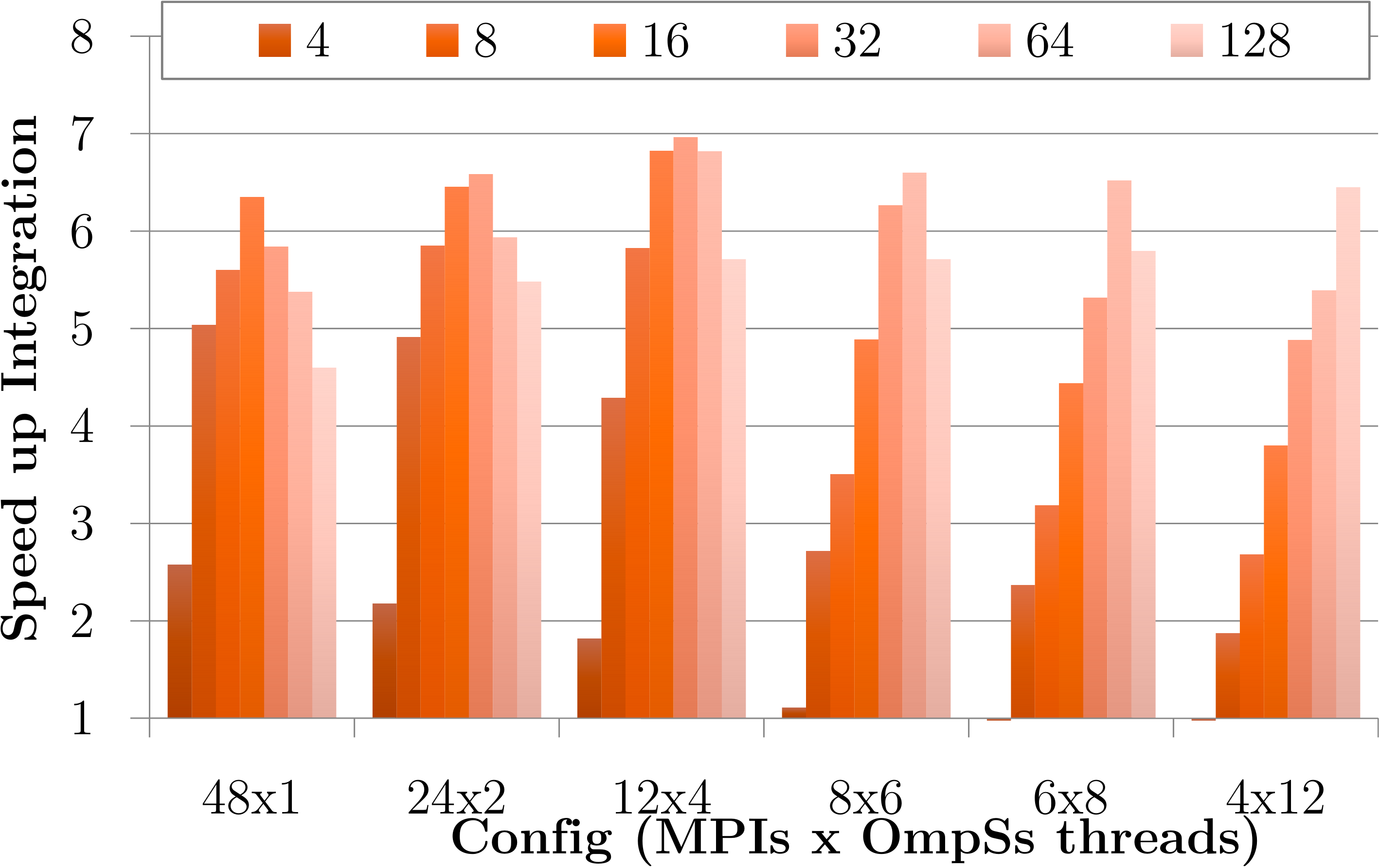}
  \captionof{figure}{Comparison of grain size impact when using DLB for reduced chemistry integration (case CF2) for different number of threads.}
  \label{fig:1node_DLB_grainsize_red}
\end{figure}

 The same study was done for the reduced chemistry case CF2 and it is shown in Figure~\ref{fig:1node_DLB_grainsize_red} where we observe more variability with respect to the grain size. Note that, since the average chemistry integration cost per element is much lower, the overhead becomes significant for low grain sizes. Consequently, for all configurations the optimal granularity is always equal to or greater than 32. Considering the optimal grain size for each configuration, the speedup obtained ranges between $6.4\times$ and $7 \times$ achieving $91\%$ of the maximum speedup for the $48\times 1$ configuration.

Regarding the whole time-step, the implementation of Alya is not completely hybrid what makes more adequate the configuration of $48\times1$ for production simulations since otherwise, other parts of the time-step would be penalized because some CPU-cores reserved to OmpSs threads would not be used. For the detailed chemistry model CF1, its integration represents $59\%$ of the overall time step, so the $2.4\times$ acceleration achieved with the $48 \times 1$ configuration results in a $1.6\times$ overall acceleration. In the reduced chemistry scenario CF2, its share is $26\%$ of the time-step, therefore its $7 \times$ acceleration results in a $1.4\times$ overall acceleration.

In this section, we have shown that DLB can improve the performance of the chemical integration loop further than the hybridization of the code. Moreover, it is confirmed that the speedup achieved by DLB depends on the imbalance present in the original execution achieving a $7\times$ speedup for highly imbalanced runs with the same number of resources.
It is seen that the selection of the grain size has an impact on the performance of DLB. Finally, large grain sizes offer less flexibility to DLB to load balance, so the best trade-off between flexibility and overhead is around a grain size of 32 for all the cases.

\subsection{Multi-node testing}
\label{}

As explained in Section~\ref{sec:dlb}, DLB can only load balance within a computational node with shared memory, so in this section we demonstrate how DLB can improve the performance of multi-node executions even though it is only acting locally inside the node. 

When considering multi-node simulations it is important to notice that, by default, resource managers spawn the MPI processes contiguously in the different computational nodes. With the continuous binding, subdomains associated with processes running in the same node tend to be adjacent in the domain. With the Round Robin (RR) binding, this locality is avoided on purpose to make each node to have different parts of the domain. This strategy is well suited for combustion simulations, since chemical reactions usually occur in specific locations of the domain and along thin layers, so the MPI ranks containing the reacting layers tend to be close to each other. In order to distribute the most loaded processes among the different nodes and improve the performance of the load balancing mechanism of DLB, it is usually better to use a Round Robin (RR) distribution of MPI processes.

\begin{figure}[htbp!]
\centering
  \includegraphics[width=\linewidth]{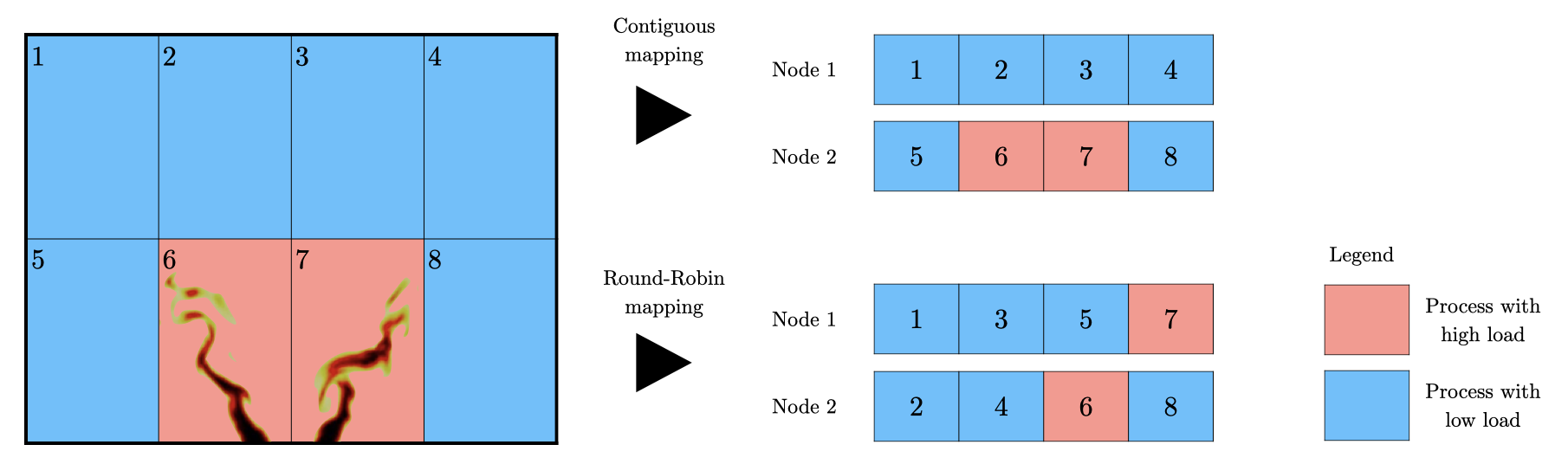}
  \captionof{figure}{Contiguous vs. Round Robin distribution of MPI ranks.}
  \label{fig:roundrobin}
\end{figure}

In Figure~\ref{fig:roundrobin}, we show an example of a combustion domain (left hand side) partitioned between 8 MPI ranks. This partition assigns the nodes of the reacting layer, which is the most computationally expensive to MPI rank 6 and 7. When a contiguous distribution of MPI ranks among nodes is done, the situation depicted in the top right hand side part of the figure is found. Where MPI ranks 1, 2, 3 and 4 are assigned to Node 1 and MPI, ranks 5, 6, 7 and 8 are assigned to Node 2. With this distribution, the two more loaded process are assigned to Node 2, producing a load imbalance across nodes that can not be addressed by DLB. However, considering a Round Robin distribution instead, bottom right hand side, Node 1 gets MPI ranks 1, 3, 5 and 7, and Node 2 gets MPI ranks 2, 4, 6 and 8. A good load balance across nodes is found with this distribution, so the load imbalance inside the nodes can be addressed with DLB.

This aspect is investigated here by the analysis of a counterflow diffusion flame, corresponding to cases CF3 and CF4 with detailed and reduced chemistry, respectively. These cases differ from the previous cases CF1 and CF2 by having a finer mesh with about 4 times larger computational load and then it will be extended to turbulent premixed flames, cases SB1 and SB2.
For these cases, we consider the range from 1 node (48 CPU-cores) up to 16 nodes (768 CPU-cores). Note that the chemistry integration does not require any MPI communication or synchronization operation and, therefore, the most limiting factor for scalability is the imbalance. All the executions in this section are done using a configuration of $48\times1$ of MPI processes and OmpSs threads per node, as we want to mimic a production run of Alya and a grain size of 32, because it is the optimum value determined in the previous section.

\begin{figure}[htbp!]
\centering
  \includegraphics[width=0.7\linewidth]{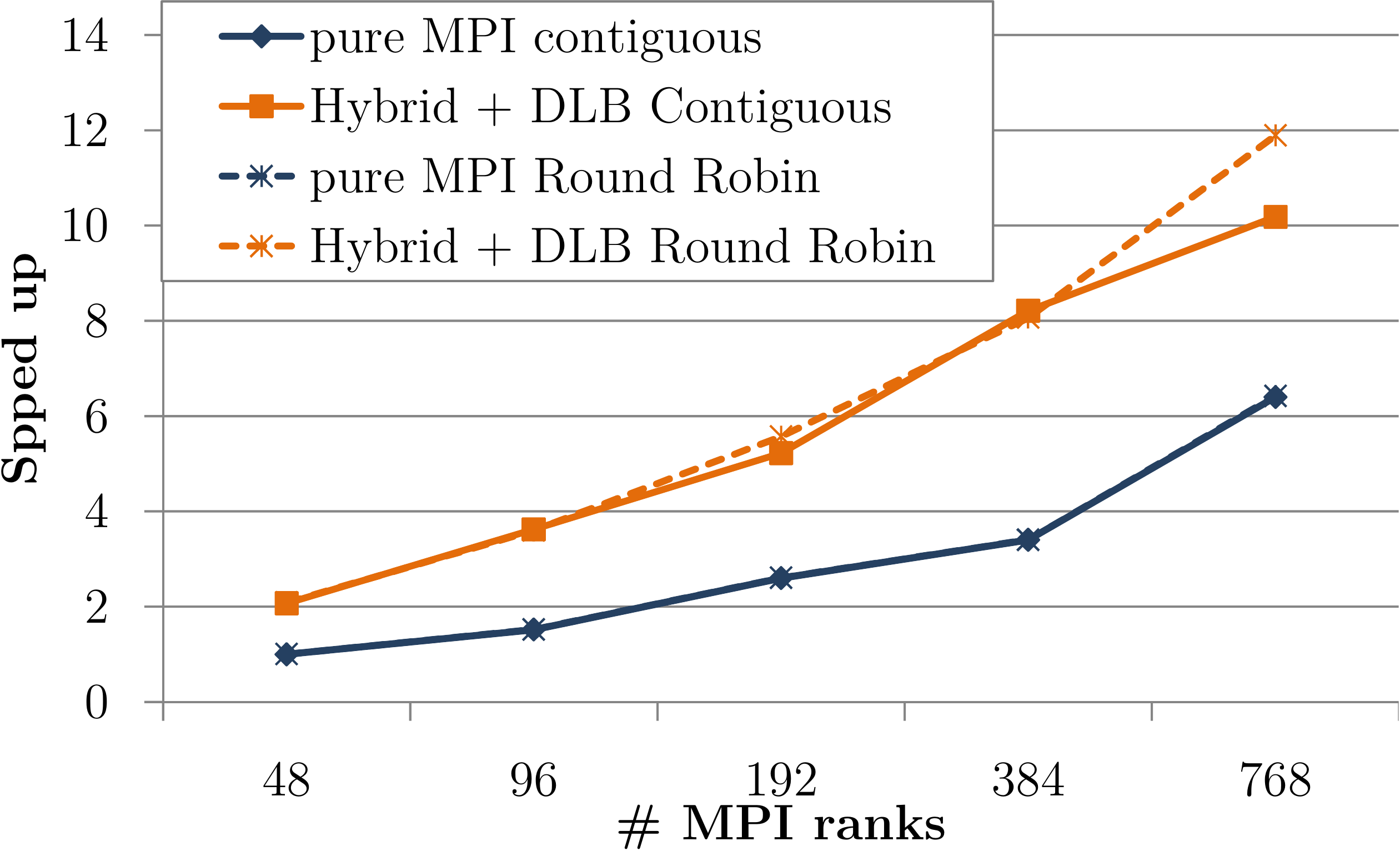}
  \captionof{figure}{Speedup-up of detailed chemical integration (case CF3) up to 16 nodes using DLB and varying distribution of MPI ranks among nodes, with a configuration of $48\times1$ and grain size 32.}
  \label{fig:scalingDetailed}
\end{figure}

In Figure~\ref{fig:scalingDetailed}, we see the speedup (Y axis) of the chemistry integration stage normalized by the MPI-only execution on 48 cores as function of the number of MPI ranks used in the simulation. In these plots, solid lines represent contiguous binding, for which MPI ranks are placed contiguously in the nodes, while dashed lines are used to represent the Round Robin binding, where MPI ranks are spawned in a Round Robin mode among the compute nodes. MPI only executions are represented with blue lines, while hybrid DLB executions are represented with orange lines.

We can see that the binding has low impact on the performance of the pure MPI implementation (blue lines overlap). Contrarily, when DLB is used, the RR binding is helpful to break the subdomain's locality and avoid situations where the subdomains associated with processes of a node cover regions with similar conditioning. In this case, we can see that for 2 nodes (96 MPI ranks) there is almost no difference. But for higher number of nodes DLB is able to obtain a better speedup when using a Round Robin distribution. With 768 MPI ranks, it obtains a $12\times$ speedup with Round Robin versus a $10\times$ speedup with contiguous distribution. Additionally, DLB improves the performance of the simulation by a factor of $2\times$ with respect to the original pure MPI run when using 16 nodes.
 
 \begin{figure}[htbp!]
\centering
  \includegraphics[width=0.7\linewidth]{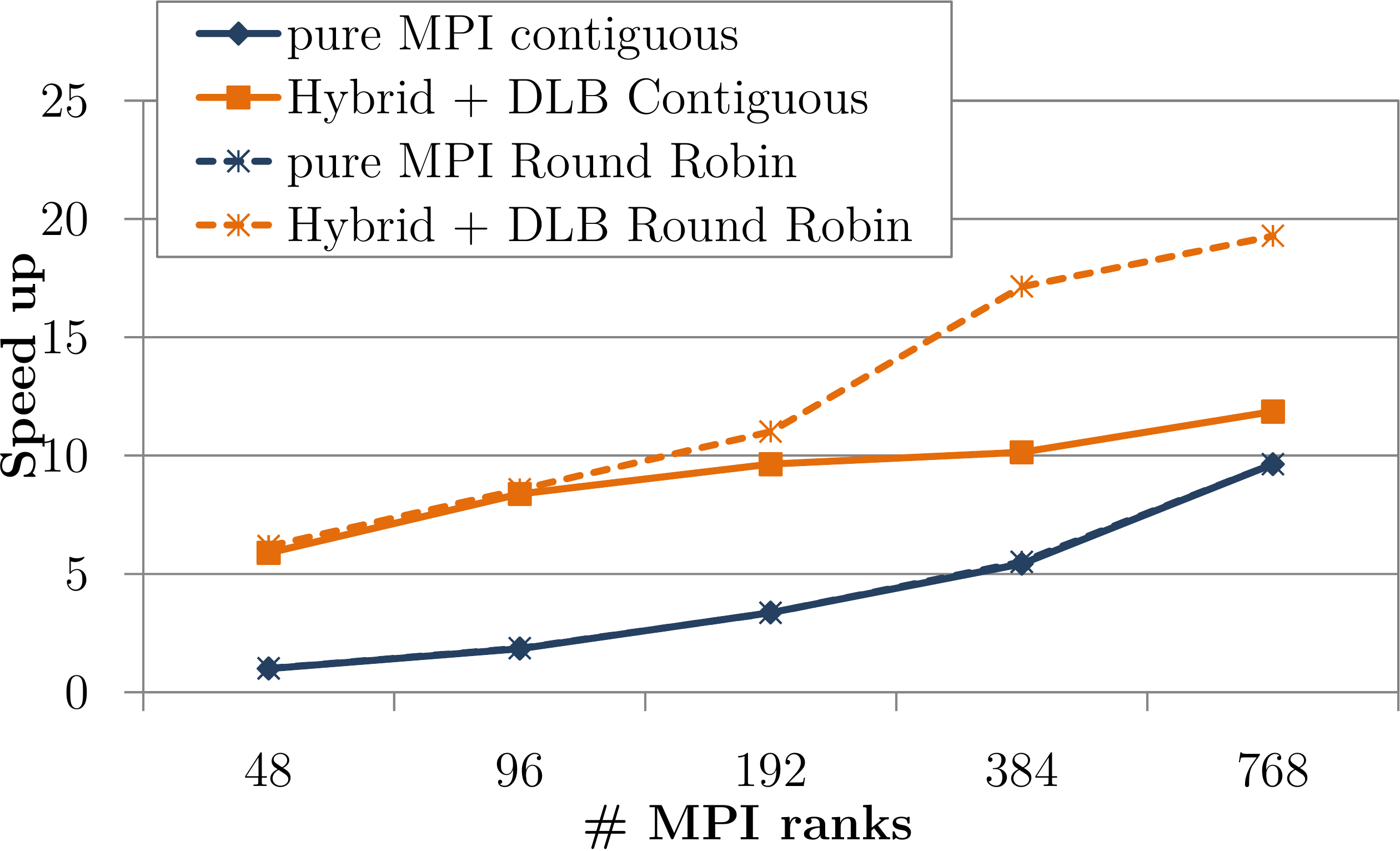}
  \captionof{figure}{Speedup-up of reduced chemical integration (case CF4) up to 16 nodes using DLB and varying distribution of MPI ranks among nodes, with a configuration of $48\times1$ and grain size 32.}
  \label{fig:scalingReduced}
\end{figure}

In Figure~\ref{fig:scalingReduced}, we see the same speedup, but for the reduced chemistry case CF4. 
The blue lines correspond to runs of the original MPI-pure code, while orange lines correspond to hybrid runs with DLB. It is observed that the distribution of MPI processes among nodes does not have an impact on the performance of the original pure MPI code (blue lines overlap), as it also occurs with the detailed chemistry case. However, we observe that the impact of the round Robin distribution in this case is even higher than for the detailed chemistry. This is due to the fact that the load was more localized in this use case, therefore, DLB benefits of distributing the most loaded processes among the different nodes 
. DLB achieves a speedup of $19\times$ when running in 16 nodes (768 MPI ranks) with Round Robin, compared to the $12\times$ speedup achieved with DLB with contiguous MPI distribution and $9\times$ with the original pure MPI code. 
 
 Summarizing, on the one hand, the scalability results for the detailed chemistry case CF3 achieve a speedup for the MPI-only version that ranges from $1\times$ (reference point) to $6\times$, while the DLB version with the RR binding ranges from $2\times$ to $12\times$. Therefore the relative scalability of each approach is the same, but being the DLB option twice as fast. On the other hand, in the reduced chemistry case CF4, the MPI-only version speedup ranges from $1\times$ to $10\times$ and the  DLB version from $6\times$ to $20\times$. In this case, the relative acceleration of the pure-MPI version is higher: the DLB version starts being six times faster and ends up being twice faster. However, still the DLB implementation clearly outperforms the pure MPI version even with very low loads per CPU-core. These results demonstrate that DLB can improve the performance of multi node runs with both detailed and reduced chemistry.
 
\subsection{Scalability study}

Finally, we want to demonstrate the use of DLB in a production run. For this, a three-dimensional turbulent premixed flame, referred here as SB1 and SB2 with detailed and reduced chemistry respectively, has been considered (see Table~\ref{tab:cases}). The scaling test is performed using from 10 nodes (480 CPU-cores) up to 40 nodes (1920 CPU-cores). This is a larger case, where the chemical reactions are more localized on specific zones of the domain and the flame features unsteady effects and fluctuations in heat release.

\begin{figure}[htbp!]
\centering
  \includegraphics[width=0.7\linewidth]{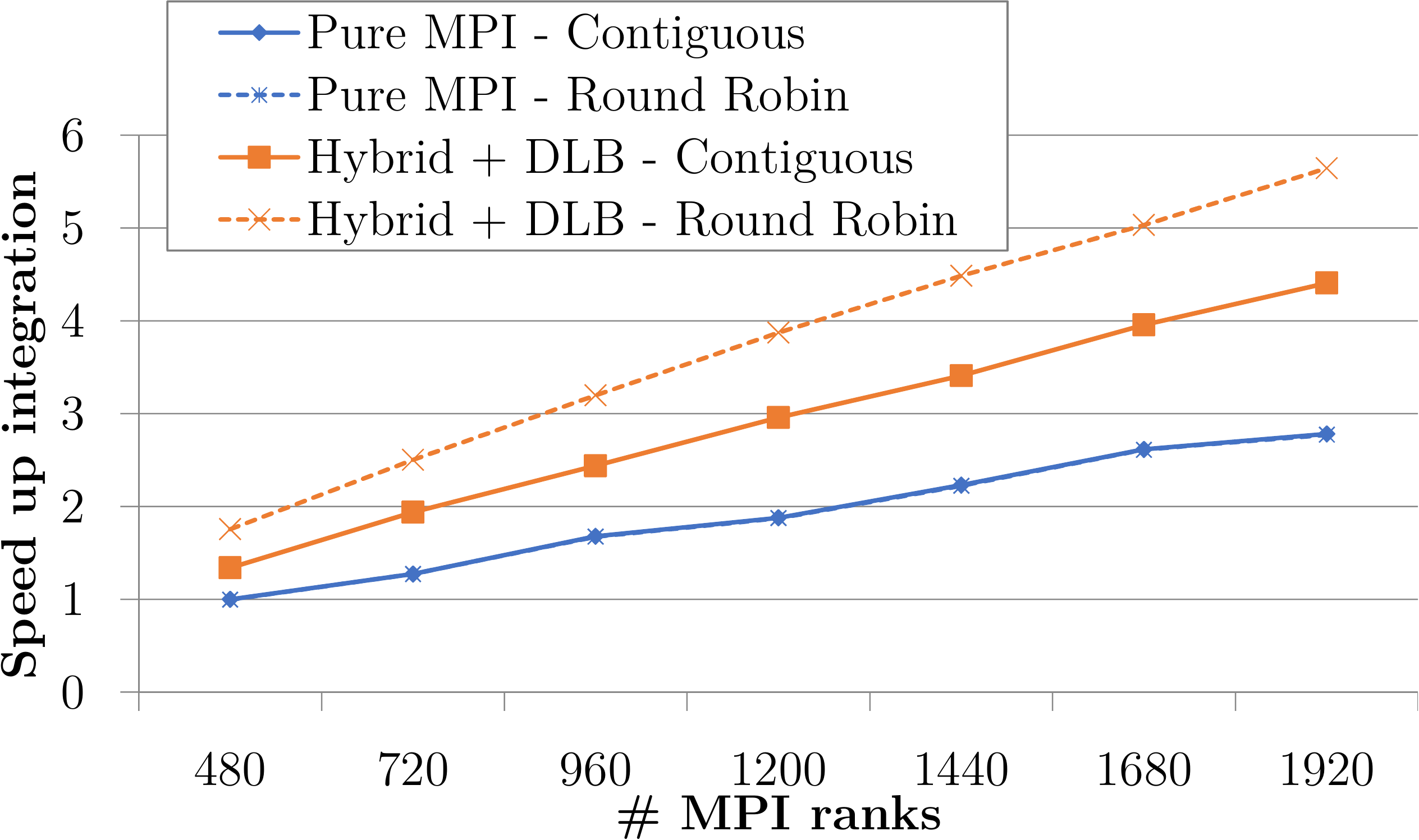}
  \captionof{figure}{Speedup-up of detailed chemical integration (case SB1) up to 40 nodes using DLB and varying distribution of MPI ranks among nodes.}
  \label{fig:scalingDetailedBigcase}
\end{figure}

In Figure~\ref{fig:scalingDetailedBigcase}, we can see the scalability study of the detailed chemistry use case SB1. A speedup with respect to the pure MPI version running on 480 cores (10 nodes) is presented up to 1920 cores. We can see how the placing of MPI processes (Contiguous or Round Robin) does not have an impact in performance when using the pure MPI version. However, when using DLB with 480 MPI ranks distributed contiguously, we obtain a speedup of $1.3\times$ while we obtain a $1.7\times$ speedup when using a Round robin distribution. More particularly, with 1920 MPI ranks the speedup obtained with DLB and a Round Robin distribution is $5.6\times$ compared to the speedup obtained with the same number of resources by the original pure MPI code of $2.7\times$. We can see that even with a large number of MPI processes distributed among several nodes, DLB can improve the performance of the execution more than $2\times$.

In fact, for the detailed case SB1, we observe three lines showing almost equal relative scalability, but at three different levels:  pure MPI as baseline,  DLB with contiguous binding $1.5 \times $ faster at each point, and  DLB with RR binding $2 \times$ faster.

\begin{figure}[htbp!]
\centering
  \includegraphics[width=0.7\linewidth]{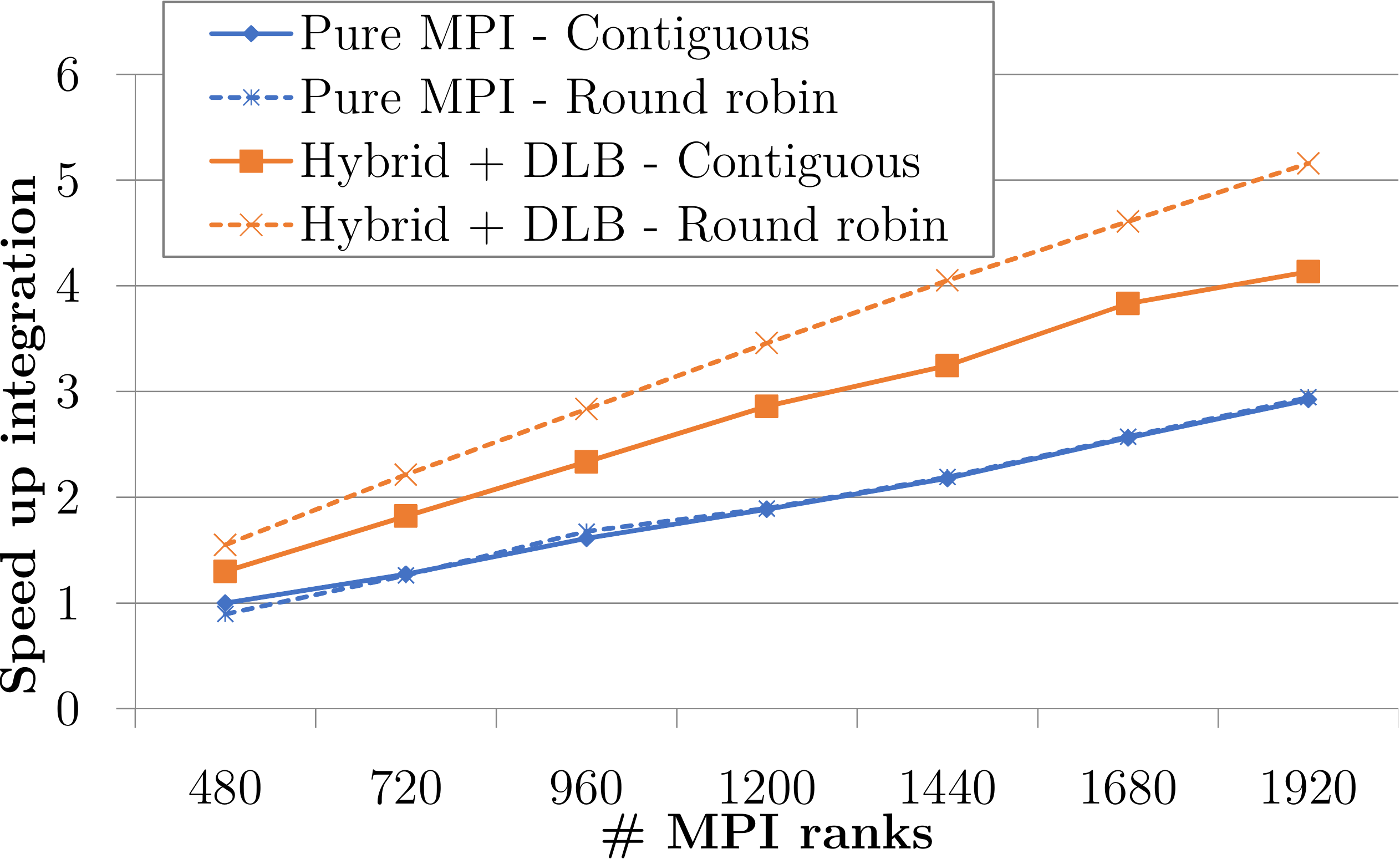}
  \captionof{figure}{Speedup of reduced chemical integration (case SB2) up to 40 nodes using DLB and varying distribution of MPI ranks among nodes.}
  \label{fig:scalingReducedBigcase}
\end{figure}

In Figure~\ref{fig:scalingReducedBigcase}, the same experiments using the reduced chemistry case SB2 are presented. Again DLB improves the performance in all the cases, obtaining even better results when using a Round Robin distribution in a similar manner than in the non-premixed flame CF4. With this use case the speedup obtained by DLB when using 1920 MPI ranks is $5.1\times$ while with the original code the speedup remained below $3\times$.

We observe that for both detailed chemistry SB1 (Figure~\ref{fig:scalingDetailedBigcase}) and reduced chemistry SB2 (Figure~\ref{fig:scalingReducedBigcase}), the performance of the  MPI-only version of the code is independent of the binding, while for the DLB version, the RR binding outperforms the continuous binding in all tests due to the joined effect of the redistribution of load in nodes with remain idle and the application of DLB.

\section{Conclusions}
\label{}

This paper presents a load balancing strategy for reaction rate evaluation and chemistry integration in reacting flow simulations. The large disparity in scales during the fuel oxidation introduces stiffness in the numerical integration of the PDEs at specific zones of the domain and this generates load imbalance in the parallel execution. This is a well known problem of the simulation of reacting flows that has been handled in various ways, primarily using inter-process workload redistribution through MPI message passing. 

This paper proposes utilizing the DLB library that allows the redistribution of computing resources at node level, lending additional CPU-cores to higher loaded MPI processes. DLB requires only two actions: i) calling a barrier function of its API, and ii) a directives-based shared memory parallelization of the region of the code under consideration. At runtime, “idle” CPU-cores are summed to the execution of higher loaded MPI processes. This strategy thus does not require explicit data transfer between MPI-processes and is activated as an automatic runtime mechanism. Furthermore, DLB is complementary to others LB mechanisms. 

In this paper, we have first implemented a task-based shared memory parallelization of the chemistry integration using OmpSs. We have studied both the sensibility to the grain size of the taskification, and the configuration of MPI and OmpSs threads within a node. A counterflow diffusion flame has been considered, using both a detailed and a reduced chemistry integration. The shared memory implementation improves the load balancing, because tasks are dynamically scheduled to the various threads running in the same MPI process, being the speedups obtained versus the pure MPI implementation, for the detailed and reduced integrations of 1.6x and 3.4x, respectively. When DLB is enabled as an additional inter process mechanism, the speedups obtained raise up to 2.3x and 7x, respectively.

Results on multi node runs show also that DLB improves the  performance of the pure-MPI code in similar proportion as single node runs. Moreover, similar scalability is obtained with the DLB implementation but on faster executions. To enhance the performance of multi-node executions an optimized distribution of the MPI processes across nodes based on a Round Robing ordering has been presented. In a production simulation of a turbulent premixed flame using 1920 CPU-cores, the enhanced DLB execution has shown to outperform by 2x the pure MPI implementation.

In consequence, the use of DLB arises as a powerful strategy to accelerate the calculations and reduce the dependency of the performance on the parallelization parameters. This becomes especially relevant in problems where there are heavy computations concentrated in specific regions of the domain and only depends on the local conditions, as it happens for the chemistry integration in reacting flows. Moreover, the use of DLB can be easily extended to other combustion models such as CMC, EDC or TPDF in which chemistry is integrated \emph{in situ}.

\section{Acknowledgements}
This work has been supported by the Spanish Government (PID2019-107255GB), by Generalitat de Catalunya (contract 2014-SGR-1051), and by the European POP CoE (GA No 824080) and EXCELLERAT project GA No 823691.

\section{Declaration of Competing Interest}
The authors declare that they have no known competing financial interests or personal relationships that could have appeared to influence the work reported in this paper.







\section{References}
\bibspacing=\dimen 100
\bibliographystyle{plain}
\bibliography{mybiblio}





\end{document}